\documentclass[12pt]{article}
\usepackage{epsfig}
\usepackage{axodraw}
\newcommand{\mysection}{\setcounter{equation}{0}\section}
\renewcommand{\theequation}{\thesection.\arabic{equation}}

\def\beq{\begin{equation}}
\def\eeq{\end{equation}}
\def\beqa{\begin{eqnarray}}
\def\eeqa{\end{eqnarray}}

\begin{document}

\begin{center}
{\Large \bf Single top quark production at the Tevatron: threshold resummation 
and finite-order soft gluon corrections}
\end{center}
\vspace{2mm}
\begin{center}
{\large Nikolaos Kidonakis}\\
\vspace{2mm}
{\it Kennesaw State University, Physics \#1202,\\
1000 Chastain Rd., Kennesaw, GA 30144-5591}\\
\end{center}

\begin{abstract}
I present a calculation of threshold soft-gluon corrections to single  
top quark production in $p{\bar p}$ collisions via all partonic processes 
in the $t$ and $s$ channels and via associated top quark and 
$W$ boson production.
The soft-gluon corrections are formally resummed to all orders, and 
finite-order expansions of the resummed cross section are calculated 
through next-to-next-to-next-to-leading order (NNNLO) at 
next-to-leading logarithmic (NLL) accuracy.
Numerical results for single top quark production at the Tevatron are 
presented, including the dependence of the cross sections on 
the top quark mass and on the factorization and renormalization scales. 
The threshold corrections in the $t$ channel are small while in the $s$ 
channel they are large and dominant. Associated $tW$ production remains 
relatively minor due to the small leading-order cross section even though 
the $K$ factors are large. 

\end{abstract}

\mysection{Introduction}

The top quark was discovered in 1995 by the CDF and D0 collaborations 
at the Tevatron in top-antitop quark pair production events \cite{CDFtt,D0tt}.
The search for single top quark events continues intensively at the 
Tevatron \cite{CDF,D0}.
The theoretical cross section is less than that of the $t{\bar t}$ cross 
section and the backgrounds make the extraction of the signal challenging.  

Single top quark production provides opportunities for the study of the 
electroweak properties of the top quark (such as a direct measurement of the 
$V_{tb}$ CKM matrix element), for further insights into electroweak
theory since the top quark mass is of the same order of magnitude 
as the electroweak 
symmetry breaking scale, as well as for the possible discovery of new physics 
(extra quarks or gauge bosons, modified top quark interactions, etc.) 
(see, for example, Refs. [5-11]). 

The production of single top quarks can proceed through three distinct 
partonic processes. One of them is the $t$-channel process that proceeds via
the exchange of a space-like $W$ boson (Fig. 1), a second is the $s$-channel 
process that proceeds via the exchange of a time-like $W$ boson (Fig. 2), 
and a third is associated $tW$ production (Fig. 3). 
In the $t$ channel we have processes of the form $qb \rightarrow q' t$ 
and ${\bar q} b \rightarrow {\bar q}' t$, such as
$ub \rightarrow dt$ and ${\bar d} b \rightarrow {\bar u} t$ as well 
as processes involving the charm quark and Cabibbo-supressed 
contributions.
In the $s$ channel we have processes of the form $q{\bar q}' \rightarrow 
{\bar b} t$, such as $u {\bar d} \rightarrow {\bar b} t$   
as well as processes involving the charm quark and Cabibbo-supressed 
contributions.
Associated production of a top quark and a $W$ boson proceeds via 
$bg \rightarrow tW^-$ as well as Cabibbo-supressed contributions.

At the Tevatron the $t$-channel process is numerically dominant, 
the $s$-channel process is smaller, and associated $tW$ production is quite 
minor.
Although we will be discussing the production of a top quark in this paper 
we note that the corresponding cross sections for the production of an antitop 
quark at the Tevatron are identical.
Calculations of NLO corrections to single top production via the various 
partonic modes have appeared in Refs. [12-20].

The cross section for single top quark production in 
proton-antiproton collisions is 
\beqa
\sigma=\sum_f \int  dx_1 dx_2 \, \phi_{f_1/p}(x_1,\mu_F)\, 
\phi_{f_2/{\bar p}}(x_2,\mu_F)\, {\hat \sigma}(s,t,u,\mu_F,\mu_R,\alpha_s) \, ,
\label{factcs}
\eeqa
where $\phi_{f_1/p}$ ($\phi_{f_2/{\bar p}}$) is the distribution 
function for parton $f_1$  ($f_2$) carrying momentum
fraction $x_1$ ($x_2$) of the proton (antiproton), $\mu_F$ is the 
factorization scale, and $\mu_R$ is the renormalization scale.
The parton-level cross section is denoted by
${\hat \sigma}$ and can be written as a perturbative expansion in the 
strong coupling $\alpha_s$, and $s$, $t$, $u$ are standard kinematical
invariants formed from the momenta of the particles in the hard scattering.

Near kinematical threshold for the production of a specified final state, 
such as a single top quark,   
large corrections appear from soft-gluon emission \cite{KS,KOS,LOS,NKtop}. 
These corrections arise from incomplete cancellations of infrared 
divergences between virtual diagrams and real diagrams with soft 
(i.e. low-energy) gluons.
In principle one can obtain the form of these soft radiative corrections 
at any order in $\alpha_s$ and formally resum them to all orders.
However in practice such resummed cross sections depend on a prescription 
to avoid the infrared singularity and ambiguities from prescription 
dependence can actually be larger than contributions from high-order terms 
\cite{NKtop}. Hence, we provide fixed-order expansions of the resummed 
cross section, as has been done for many other processes 
\cite{NKAB,NKtop,NKJO,KVtbc,NKASV,NKch,NKuni}, in order to avoid 
such ambiguities. 
We will calculate soft-gluon corrections for single top quark production  
through NNNLO at NLL accuracy.

\begin{figure}[htb]
\begin{center}
\begin{picture}(120,120)(0,0)
\Line(0,80)(60,70)
\Line(60,70)(120,80)
\Line(0,20)(60,30)
\Line(60,30)(120,20)
\Photon(60,70)(60,30){3}{5}
\Text(0,10)[c]{$b$}\Text(0,90)[c]{$q$ (${\bar q}$)}
\Text(75,50)[c]{$W$}
\Text(120,10)[c]{$t$}\Text(120,90)[c]{$q'$ (${\bar q}'$)}
\end{picture}
\end{center}
\vspace{-5mm}
\caption{\label{tlo} Leading-order $t$-channel diagram for 
single top quark production.}
\end{figure}
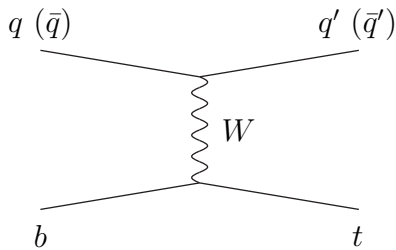

\begin{figure}[htb]
\begin{center}
\begin{picture}(120,120)(0,0)
\Line(0,75)(30,50)
\Line(0,25)(30,50)
\Line(90,50)(120,75)
\Line(90,50)(120,25)
\Photon(30,50)(90,50){3}{5}
\Text(0,15)[c]{${\bar q'}$}\Text(0,85)[c]{$q$}
\Text(60,65)[c]{$W$}
\Text(120,15)[c]{$t$}\Text(120,85)[c]{${\bar b}$}
\end{picture}
\end{center}
\vspace{-5mm}
\caption{\label{slo} Leading-order $s$-channel diagram for 
single top quark production.}
\end{figure}
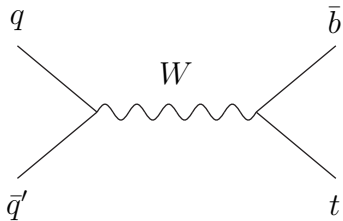

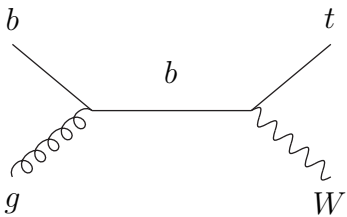
\begin{figure}[htb]
\begin{center}
\begin{picture}(120,120)(0,0)
\Line(0,75)(30,50)
\Gluon(0,25)(30,50){3}{5}
\Line(30,50)(90,50)
\Photon(90,50)(120,25){3}{5}
\Line(90,50)(120,75)
\Text(0,15)[c]{$g$}\Text(0,85)[c]{$b$}
\Text(60,65)[c]{$b$}
\Text(120,15)[c]{$W$}\Text(120,85)[c]{$t$}
\end{picture}
\end{center}
\vspace{-5mm}
\caption{\label{alo} Leading-order associated $tW$ production diagram for 
single top quark production.}
\end{figure}

In Section 2 we briefly describe the threshold resummation formalism and 
provide expressions for the resummed cross section. In Section 3 we 
expand the resummed cross section in powers of $\alpha_s$ and provide formulas 
for the soft-gluon corrections through NNNLO. 
In Section 4 we present numerical results for single top quark production via 
the $t$ channel at the Tevatron. Analogous results are provided for the 
$s$ channel in Section 5 and for associated $tW$ production in Section 6.
The conclusion is in Section 7, and the Appendix collects formulas and 
details on the kinematics and on electroweak parameters used in the 
calculations.

\mysection{Threshold resummation}

In this section we present the analytical form of the 
resummed cross section for single top quark production.  
Details of the general resummation formalism for hard-scattering cross 
sections have been presented elsewhere \cite{KS,KOS,LOS,NKtop,NKuni,NKNNNLO}
so here we explicitly show only the 
expressions directly relevant to single top quark production. 

For the process $p_1+p_2 \rightarrow p_3+p_4$ (where the particles 
are represented by their momenta $p_i$),
the partonic kinematical invariants are 
$s=(p_1+p_2)^2$, $t=(p_1-p_3)^2$, $u=(p_2-p_3)^2$, 
$s_4=s+t+u-m_3^2-m_4^2$.
Note that we ignore the mass of the $b$-quark in the kinematics.
Near threshold, i.e. when we have just enough
partonic energy to produce the final state,  $s_4 \rightarrow 0$.
The threshold corrections then take the form of logarithmic plus 
distributions, $[\ln^l(s_4/m_t^2)/s_4]_+$, 
where $l\le 2n-1$ for the $n$-th order QCD corrections and $m_t$ is 
the top quark mass. 
These plus distributions are defined 
through their integral with any smooth function, such as parton distributions, 
giving a finite result \cite{KS,NKNNNLO}. 

The resummation of threshold logarithms is carried out in moment space and it 
follows from the refactorization of the cross section into hard, soft, and 
jet functions that describe, respectively, the hard scattering, noncollinear 
soft gluon emission, and collinear gluon emission from the partons 
in the scattering \cite{KS,KOS,LOS}. 
We define moments of the partonic cross section by 
${\hat\sigma}(N)=\int (ds_4/s) \;  e^{-N s_4/s} {\hat\sigma}(s_4)$,
with $N$ the moment variable. Under moments logarithms of $s_4$ transform 
into logarithms of $N$, which exponentiate.
The resummed partonic cross section in moment space \cite{KS,KOS,LOS,NKNNNLO} 
is then given by  
\beqa
{\hat{\sigma}}^{res}(N) &=&
\exp\left[ \sum_i E^{f_i}(N_i)\right] \;
\exp\left[ \sum_j {E'}^{f_j}(N_j)\right] \;
\exp \left[\sum_i 2\int_{\mu_F}^{\sqrt{s}} \frac{d\mu}{\mu}\;
\gamma_{i/i}\left(N_i,\alpha_s(\mu)\right)\right] \;
\nonumber\\ && \hspace{-10mm} \times \,
\exp \left[\sum_i 2 d_{\alpha_s} \int_{\mu_R}^{\sqrt{s}} \frac{d\mu}{\mu}\;
\beta\left(\alpha_s(\mu)\right)\right] \;
H^{f_i f_j}\left(\alpha_s(\mu_R)\right) 
\nonumber\\ && \hspace{-10mm} \times \,
\exp \left[\int_{\sqrt{s}}^{{\sqrt{s}}/{\tilde N_j}}
\frac{d\mu}{\mu} \;
\Gamma_S^{\dagger \, f_i f_j}\left(\alpha_s(\mu)\right)\right] \;
{\tilde S^{f_i f_j}} \left(\alpha_s\left(\frac{\sqrt{s}}{\tilde N_j}\right) 
\right) \;
\exp \left[\int_{\sqrt{s}}^{{\sqrt{s}}/{\tilde N_j}}
\frac{d\mu}{\mu}\; \Gamma_S^{f_i f_j}
\left(\alpha_s(\mu)\right)\right] \, .
\nonumber \\ 
\label{resHS}
\eeqa

The first exponent resums collinear and soft gluon emission 
from the incoming partons in the hard scattering and 
is given in the $\overline{\rm MS}$ scheme by
\beq
\sum_i E^{f_i}(N_i)=
-C_i \sum_i \int^1_0 dz \frac{z^{N_i-1}-1}{1-z}\;
\left \{\int^1_{(1-z)^2} \frac{d\lambda}{\lambda}
\frac{\alpha_s(\lambda s)}{\pi}
+\frac{\alpha_s((1-z)^2 s)}{\pi}\right\}+{\cal O}(\alpha_s^2) \, .
\label{Eexp}
\eeq
Here $N_1=N [(-u+m_t^2)/m_t^2]$  and 
$N_2=N [(-t+m_t^2)/m_t^2]$ for incoming partons $i=1,2$,  
$C_i=C_F=(N_c^2-1)/(2N_c)$ for quarks, with $N_c$ the number of colors,
and $C_i=C_A=N_c$ for gluons.

The second exponent resums collinear and soft gluon emission 
from the outgoing massless partons, if any (none in associated 
production), in the hard scattering and 
is given in the $\overline{\rm MS}$ scheme by
\beqa
{E'}^{f_j}(N_j)&\! \!=&\! \!
\int^1_0 dz \frac{z^{N_j-1}-1}{1-z}\;
\left \{C_j \int^{1-z}_{(1-z)^2} \frac{d\lambda}{\lambda}
\frac{\alpha_s\left(\lambda s\right)}{\pi}
-B_j^{(1)}\frac{\alpha_s((1-z)s)}{\pi}
-C_j\frac{\alpha_s((1-z)^2 s)}{\pi}\right\}
\nonumber \\ &&
+{\cal O}(\alpha_s^2) \, .
\label{Ejexp}
\eeqa
Here $N_j=N (s/m_t^2)$, and 
$B_j^{(1)}$ equals $3C_F/4$ for quarks and $\beta_0/4$ for gluons, 
with $\beta_0=(11C_A-2n_f)/3$ the lowest-order term in the expansion 
of the $\beta$-function, $\beta(\alpha_s)\equiv \mu d\ln g/d\mu=
-\beta_0 \alpha_s/(4\pi)+{\cal O}(\alpha_s^2)$, where $n_f$ is the number of 
light quark flavors.

In the third exponent, the parton anomalous dimensions are 
given by $\gamma_{q/q}(N_i,\alpha_s)=$ 
\newline $(\alpha_s/\pi) [3C_F/4-C_F \ln N_i]
+{\cal O}(\alpha_s^2)$ for quarks and 
$\gamma_{g/g}(N_i,\alpha_s)=(\alpha_s/\pi) [\beta_0/4-C_A \ln N_i]
+{\cal O}(\alpha_s^2)$ for gluons, and $\mu_F$ is the factorization scale.

In the fourth exponent the $\beta$ function is as described above,   
the constant $d_{\alpha_s}=0$ or 1 if the leading order cross section is 
of order $\alpha_s^0$ ($t$ and $s$ channels) or $\alpha_s^1$ (associated
$tW$ production) respectively, and $\mu_R$ is the renormalization scale.  

$H^{f_if_j}$ are the hard-scattering functions
for the scattering of partons $f_i$ and $f_j$, while $S^{f_if_j}$ are the
soft functions describing noncollinear soft gluon emission \cite{KS}.
We use the expansions
$H=H^{(0)}+(\alpha_s/\pi)H^{(1)}+{\cal O}(\alpha_s^2)$
 and
$S=S^{(0)}+(\alpha_s/\pi)S^{(1)}+{\cal O}(\alpha_s^2)$.
Also ${\tilde N}=N e^{\gamma_E}$, with $\gamma_E$ the Euler constant.
At lowest order, the product of $H^{f_if_j}$ and $S^{f_if_j}$
reproduces the Born cross section for each partonic process,
$\sigma^B=\alpha_s^{d_{\alpha_s}}H^{(0)}S^{(0)}$.

The evolution of the soft function
follows from its renormalization group properties and
is given in terms of the soft anomalous dimension $\Gamma_S^{f_if_j}$
\cite{KS,KOS,ADS}. We expand 
$\Gamma_S=(\alpha_s/\pi)\Gamma_S^{(1)}+{\cal O}(\alpha_s^2)$.
The one-loop term $\Gamma_S^{(1)}$ is determined through the explicit 
calculation, in the eikonal approximation, of the one-loop diagrams 
involving  eikonal vertex corrections as well as top quark self-energies
as shown in Figs. 4-7 (for the requisite integrals in dimensional 
regularization see \cite{KS}). Here the partons in the scattering  
are represented by eikonal lines that connect in a color vertex.
Explicit expressions for the soft anomalous dimensions in the various 
channels are given in the next section.

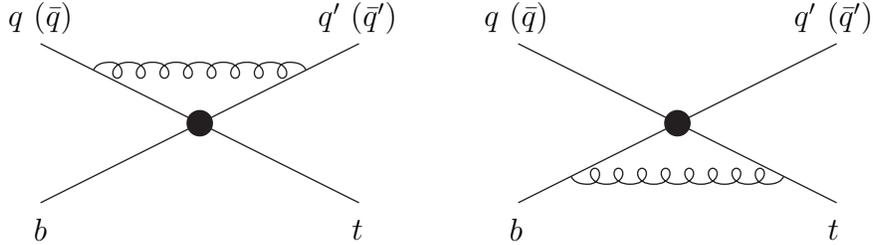
\begin{figure}[htb]
\begin{center}
\begin{picture}(300,120)(0,0)
\Line(0,80)(60,50)
\Line(60,50)(120,80)
\Line(0,20)(60,50)
\Line(60,50)(120,20)
\Gluon(20,70)(100,70){3}{8}
\Vertex(60,50){5} 
\Text(0,10)[c]{$b$}\Text(0,90)[c]{$q$ (${\bar q}$)}
\Text(120,10)[c]{$t$}\Text(120,90)[c]{$q'$ (${\bar q}'$)}
\Line(180,80)(240,50)
\Line(240,50)(300,80)
\Line(180,20)(240,50)
\Line(240,50)(300,20)
\Gluon(280,30)(200,30){3}{8}
\Vertex(240,50){5} 
\Text(180,10)[c]{$b$}\Text(180,90)[c]{$q$ (${\bar q}$)}
\Text(300,10)[c]{$t$}\Text(300,90)[c]{$q'$ (${\bar q}'$)}
\end{picture}
\end{center}
\vspace{-5mm}
\caption{\label{t1l} One-loop eikonal vertex corrections to the soft function 
for the $t$-channel diagram in single top quark production.}
\end{figure}

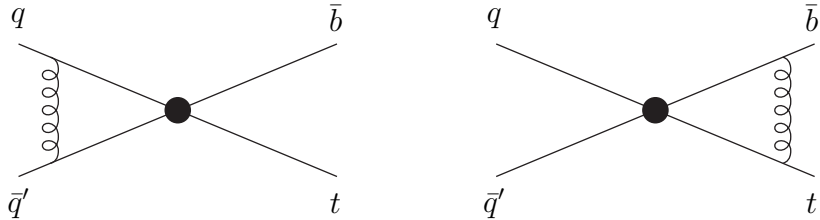
\begin{figure}[htb]
\begin{center}
\begin{picture}(300,120)(0,0)
\Line(0,75)(60,50)
\Line(0,25)(60,50)
\Line(60,50)(120,75)
\Line(60,50)(120,25)
\Gluon(12,70)(12,30){3}{5}
\Vertex(60,50){5}
\Text(0,15)[c]{${\bar q'}$}\Text(0,85)[c]{$q$}
\Text(120,15)[c]{$t$}\Text(120,85)[c]{${\bar b}$}
\Line(180,75)(240,50)
\Line(180,25)(240,50)
\Line(240,50)(300,75)
\Line(240,50)(300,25)
\Gluon(288,70)(288,30){3}{5}
\Vertex(240,50){5}
\Text(180,15)[c]{${\bar q'}$}\Text(180,85)[c]{$q$}
\Text(300,15)[c]{$t$}\Text(300,85)[c]{${\bar b}$}
\end{picture}
\end{center}
\vspace{-5mm}
\caption{\label{s1l} One-loop eikonal vertex corrections to the soft function 
for the $s$-channel diagram in single top quark production.} 
\end{figure}

\begin{figure}[htb]
\begin{center}
\begin{picture}(400,120)(0,0)
\Line(0,75)(50,50)
\Gluon(0,25)(50,50){3}{5}
\Line(50,50)(100,50)
\Vertex(50,50){5}
\Gluon(10,70)(10,33){3}{4}
\Text(0,15)[c]{$g$}\Text(0,85)[c]{$b$}
\Text(100,40)[c]{$t$}
\Line(150,75)(200,50)
\Gluon(150,25)(200,50){3}{5}
\Line(200,50)(250,50)
\Vertex(200,50){5}
\GlueArc(200,45)(25,11,144){3}{5}
\Text(150,15)[c]{$g$}\Text(150,85)[c]{$b$}
\Text(250,40)[c]{$t$}
\Line(300,75)(350,50)
\Gluon(300,25)(350,50){3}{5}
\Line(350,50)(400,50)
\Vertex(350,50){5}
\GlueArc(350,55)(25,211,349){3}{5}
\Text(300,15)[c]{$g$}\Text(300,85)[c]{$b$}
\Text(400,40)[c]{$t$}
\end{picture}
\end{center}
\vspace{-5mm}
\caption{\label{a1l} One-loop eikonal vertex corrections to the soft function 
for the associated $tW$ production diagram in single top quark production.}
\end{figure}
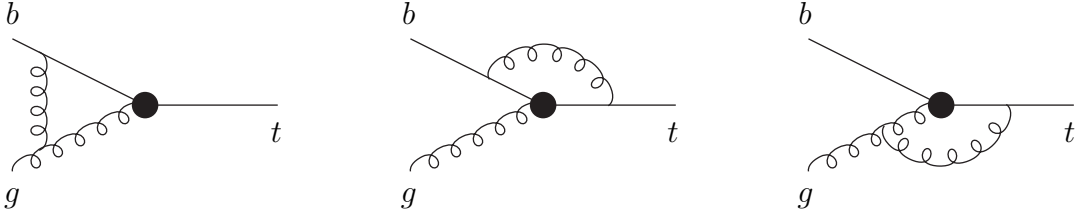

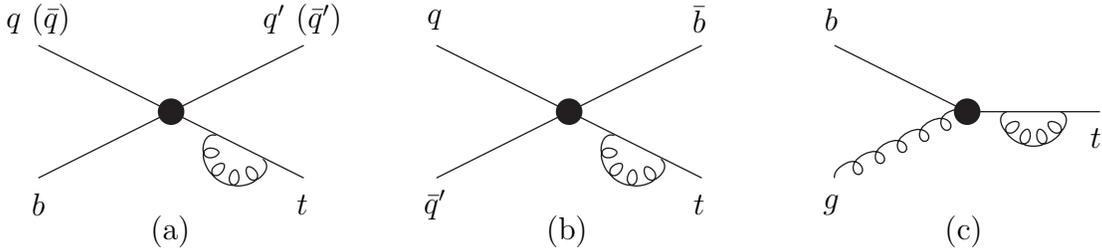
\begin{figure}[htb]
\begin{center}
\begin{picture}(400,120)(0,0)
\Line(0,75)(50,50)
\Line(50,50)(100,75)
\Line(0,25)(50,50)
\Line(50,50)(100,25)
\Vertex(50,50){5} 
\GlueArc(75,35)(10,143,343){3}{4}
\Text(0,15)[c]{$b$}\Text(0,85)[c]{$q$ (${\bar q}$)}
\Text(100,15)[c]{$t$}\Text(100,85)[c]{$q'$ (${\bar q}'$)}
\Text(50,5)[c]{(a)}
\Line(150,75)(200,50)
\Line(150,25)(200,50)
\Line(200,50)(250,75)
\Line(200,50)(250,25)
\Vertex(200,50){5}
\GlueArc(225,35)(10,143,343){3}{4}
\Text(150,15)[c]{${\bar q'}$}\Text(150,85)[c]{$q$}
\Text(250,15)[c]{$t$}\Text(250,85)[c]{${\bar b}$}
\Text(200,5)[c]{(b)}
\Line(300,75)(350,50)
\Gluon(300,25)(350,50){3}{5}
\Line(350,50)(400,50)
\Vertex(350,50){5}
\GlueArc(375,50)(10,180,360){3}{4}
\Text(300,15)[c]{$g$}\Text(300,85)[c]{$b$}
\Text(400,40)[c]{$t$}
\Text(350,5)[c]{(c)}
\end{picture}
\end{center}
\vspace{-5mm}
\caption{\label{topse} Top-quark eikonal self-energy one-loop corrections 
in single top quark production: (a) $t$ channel; 
(b) $s$ channel; (c) associated $tW$ production.}
\end{figure}

By expanding the resummed cross section, Eq. (\ref{resHS}), 
in powers of $\alpha_s$ we derive 
fixed-order corrections in the perturbative series that do not involve the 
prescription ambiguity that a fully resummed cross section entails 
\cite{NKtop}.
In the $n$-th order corrections, the leading logarithms (LL) are those 
with $l=2n-1$ while the next-to-leading logarithms (NLL) are those with 
$l=2n-2$. In this paper we calculate next-to-leading order (NLO), 
next-to-next-to-leading order (NNLO), and NNNLO soft-gluon threshold 
corrections at NLL accuracy, i.e. at each order including both leading and 
next-to-leading logarithms, and also consistently including terms
that involve the factorization and renormalization scales and  
$\zeta$ constants, which arise from the inversion from moment space  
where the resummation is performed back to momentum space. 
We denote these corrections as NLO-NLL, NNLO-NLL, and NNNLO-NLL, respectively.
Full details are given in the next section (see also Refs. 
\cite{NKtop,NKNNNLO}).
  
\mysection{NNNLO soft-gluon corrections}

We now proceed with the calculation of the NLO, NNLO, and NNNLO soft-gluon
corrections at NLL accuracy by expanding the resummed cross section, 
Eq. (\ref{resHS}). In our derivation of these corrections
we follow the general techniques and master formulas presented
in Ref. \cite{NKNNNLO}.

The NLO soft-gluon corrections at NLL accuracy for the processes are
\beq
\frac{d^2{\hat\sigma}^{(1)}}{dt \, du}
=F^B
\frac{\alpha_s(\mu_R^2)}{\pi} \left\{
c_{3} \left[\frac{\ln(s_4/m_t^2)}{s_4}\right]_+
+c_{2} \left[\frac{1}{s_4}\right]_+
+c_{1}^{\mu} \, \delta(s_4)\right\} \, ,
\label{NLO}
\eeq
where $F_B$ is the Born term defined in the Appendix, Eq. (\ref{fb}).

For the $t$ and $s$ channels the LL coefficient is $c_{3}^{t,s}=3C_F$ 
while for associated $tW$ production it is $c_3^{tW}=2(C_F+C_A)$.

The NLL coefficient, $c_2$, can be written as 
$c_2=T_2+c_2^{\mu}$, 
where $T_2$ represents the scale-independent part of $c_2$ and  
$c_2^{\mu}$ has all the scale dependence.
For the $t$ and $s$ channels,
\beq
T_{2}^{t,s}=2 {\rm Re} {\Gamma'}_S^{(1)\; t,s}
-\frac{15}{4}C_F-2C_F\ln\left(\frac{(t-m^2_t)(u-m^2_t)}{m_t^4}\right)
-3C_F\ln\left(\frac{m_t^2}{s}\right)
\eeq
and
\beq
c_2^{\mu \; t,s}= -2C_F \ln\left(\frac{\mu_F^2}{m_t^2}\right) \, ,
\eeq
while for associated $tW$ production, 
\beqa
T_{2}^{tW} &=& 2 {\rm Re} {\Gamma'}_S^{(1)\; tW}
-C_F-C_A-2C_F\ln\left(\frac{-u+m^2_W}{m_t^2}\right)
-2C_A\ln\left(\frac{-t+m^2_W}{m_t^2}\right)
\nonumber \\ && \hspace{5mm}
{}-(C_F+C_A)\ln\left(\frac{m_t^2}{s}\right),
\eeqa
with $m_W$ the $W$-boson mass,
and
\beq
c_2^{\mu \; tW}= -(C_F+C_A) \ln\left(\frac{\mu_F^2}{m_t^2}\right) \, .
\eeq
The term ${\rm Re} {\Gamma'}_S^{(1)}$ 
denotes the real gauge-independent part
of the one-loop soft anomalous dimension 
(gauge-dependent terms cancel out in the cross section). 
A one-loop calculation gives for the $t$ channel (Figs. 4, 7a)
\beq
{\rm Re}{\Gamma'}_S^{(1)\; t}
=C_F \left[\ln\left(\frac{-t}{s}\right)
+\ln\left(\frac{m_t^2-t}{m_t\sqrt{s}}\right)+1\right]\, ,
\eeq
for the $s$ channel (Figs. 5, 7b) 
\beq
{\rm Re}{\Gamma'}_S^{(1)\; s}
=C_F \left[\ln\left(\frac{s-m_t^2}{m_t\sqrt{s}}\right)+1\right]\, ,
\eeq
and for associated $tW$ production (Figs. 6, 7c)
\beq
{\rm Re}{\Gamma'}_S^{(1)\; tW}
=C_F \ln\left(\frac{m_t^2-t}{m_t\sqrt{s}}\right)
+\frac{C_A}{2} \ln\left(\frac{m_t^2-u}{m_t^2-t}\right)
+\frac{C_A}{2} \, .
\eeq

The coefficient $c_1^{\mu}$ in Eq. (\ref{NLO}) represents the 
scale-dependent part of the $\delta(s_4)$ corrections.
For the $t$ and $s$ channels
\beq
c_1^{\mu \; t,s}=\left[C_F \ln\left(\frac{(t-m^2_t)(u-m^2_t)}{m_t^4}\right)
-\frac{3}{2}C_F\right]\ln\left(\frac{\mu_F^2}{m_t^2}\right) \, ,
\eeq
and for associated $tW$ production
\beqa
c_1^{\mu \; tW}&=&\left[C_F \ln\left(\frac{-u+m^2_W}
{m_t^2}\right)
+C_A \ln\left(\frac{-t+m^2_W}{m_t^2}\right)
-\frac{3}{4}C_F-\frac{\beta_0}{4}\right]
\ln\left(\frac{\mu_F^2}{m_t^2}\right)
\nonumber \\ && \hspace{5mm}
{}+\frac{\beta_0}{4} \ln\left(\frac{\mu_R^2}{m_t^2}\right) \, .
\eeqa
Note that we do not calculate the virtual
corrections here. Our calculation of the NLO soft-gluon corrections
includes the full leading and next-to-leading logarithms and is thus
a NLO-NLL calculation. 

We next calculate the NNLO soft-gluon corrections. In the $t$ and 
$s$ channels the corrections take the form 
\beqa
&& \hspace{-5mm}\frac{d^2{\hat\sigma}^{(2)}_{t,s}}
{dt \, du}
=F^B \frac{\alpha_s^2(\mu_R^2)}{\pi^2} 
\left\{\frac{1}{2} c_3^2 
\left[\frac{\ln^3(s_4/m_t^2)}{s_4}\right]_+ 
+\left[\frac{3}{2} c_3 \, c_2
-\frac{\beta_0}{4} c_3 +C_F \frac{\beta_0}{8}\right] 
\left[\frac{\ln^2(s_4/m_t^2)}{s_4}\right]_+ \right.
\nonumber \\ && 
{}+\left[c_3 \, c_1^{\mu}
+\left(c_2^{\mu}\right)^2
+2\, c_2^{\mu} \, T_2
+\frac{\beta_0}{4} c_3 
\ln\left(\frac{\mu_R^2}{m_t^2}\right)
-\zeta_2 \, c_3^2 \right]
\left[\frac{\ln(s_4/m_t^2)}{s_4}\right]_+
\nonumber \\ &&  
{}+\left[c_2^{\mu} \, c_1^{\mu}
+\frac{\beta_0}{4}  c_2^{\mu}
\ln\left(\frac{\mu_R^2}{m^2_t}\right) 
+C_F\frac{\beta_0}{4} \ln^2\left(\frac{\mu_F^2}{m_t^2}\right)
-\zeta_2 \, c_3 \, c_2
+\zeta_3 \, c_3^2\right]
\left[\frac{1}{s_4}\right]_+  
\nonumber \\ &&  \left.
{}+\left[-\frac{\zeta_2}{2}\left((c_2^{\mu})^2+2 T_2 c_2^{\mu}\right)
+\zeta_3 c_3 c_2^{\mu} \right]
\delta(s_4) \right\} \, ,
\label{NNLOts}
\eeqa
where $\zeta_2=\pi^2/6$ and $\zeta_3=1.2020569\cdots$, 
and where for $c_3$, $c_2$, $T_2$, $c_2^{\mu}$, and $c_1^{\mu}$ we 
use the values given previously for each channel.

For associated $tW$ production the corrections are
\beqa
&& \hspace{-5mm}\frac{d^2{\hat\sigma}^{(2)}_{tW}}
{dt \, du}
=F^B \frac{\alpha_s^2(\mu_R^2)}{\pi^2} 
\left\{\frac{1}{2} c_3^2 
\left[\frac{\ln^3(s_4/m_t^2)}{s_4}\right]_+ 
+\left[\frac{3}{2} c_3 \, c_2
-\frac{\beta_0}{4} c_3 \right] 
\left[\frac{\ln^2(s_4/m_t^2)}{s_4}\right]_+ \right.
\nonumber \\ && 
{}+\left[c_3 \, c_1^{\mu}
+\left(c_2^{\mu}\right)^2
+2\, c_2^{\mu} \, T_2
+\frac{\beta_0}{4} c_3 
\ln\left(\frac{\mu_R^2}{m_t^2}\right)
-\zeta_2 \, c_3^2 \right]
\left[\frac{\ln(s_4/m_t^2)}{s_4}\right]_+
\nonumber \\ &&  
{}+\left[c_2^{\mu} \, c_1^{\mu}
+\frac{\beta_0}{4}  c_2^{\mu}
\ln\left(\frac{\mu_R^2}{m^2_t}\right) 
+(C_F+C_A)\frac{\beta_0}{8} \ln^2\left(\frac{\mu_F^2}{m_t^2}\right)
-\zeta_2 \, c_3 \, c_2
+\zeta_3 \, c_3^2\right]
\left[\frac{1}{s_4}\right]_+  
\nonumber \\ &&  \left.
{}+\left[-\frac{\zeta_2}{2}\left((c_2^{\mu})^2+2 T_2 c_2^{\mu}\right)
+\zeta_3 c_3 c_2^{\mu} \right]
\delta(s_4) \right\} \, ,
\label{NNLOtW}
\eeqa 
where for $c_3$, $c_2$, $T_2$, $c_2^{\mu}$, and $c_1^{\mu}$ we 
use the values given previously for the $tW$ channel.
We see that this is very similar to the expression for the $t$ and 
$s$ channels, Eq. (\ref{NNLOts}), differing only by some $\beta_0$ terms 
in the $[\ln^2(s_4/m_t^2)/s_4]_+$ and $[1/s_4]_+$ terms.
 
We note that only the leading and next-to-leading logarithms are complete, 
i.e. are fully known.
Hence this is a NNLO-NLL calculation.
Consistent with a NLL calculation \cite{NKtop,NKNNNLO} 
we have also kept all logarithms of the 
factorization and renormalization scales in the
$[\ln(s_4/m_t^2)/s_4]_+$ term, and squares of logarithms
of the scales in the $[1/s_4]_+$ term. 
We have also kept  $\zeta_2$ and $\zeta_3$ terms
that arise in the calculation of the soft corrections 
when inverting from moments back to momentum space \cite{NKtop,NKNNNLO}.
This includes all $\zeta$ terms in the $[\ln(s_4/m_t^2)/s_4]_+$ 
and $[1/s_4]_+$ terms, and $\zeta$ terms multiplying logarithms of the scales
in the $\delta(s_4)$ term.

At NNNLO for all channels the corrections take the form 
\beqa
\frac{d^2{\hat\sigma}^{(3)}}{dt \, du}
&=&F^B \frac{\alpha_s^3(\mu_R^2)}{\pi^3} 
\left\{\frac{1}{8} c_3^3 
\left[\frac{\ln^5(s_4/m_t^2)}{s_4}\right]_+ 
+\left[\frac{5}{8} c_3^2 \; c_2 
-\frac{5}{2} c_3 X_3\right] 
\left[\frac{\ln^4(s_4/m_t^2)}{s_4}\right]_+  \right. 
\nonumber \\ && 
{}+\left[c_3 \, (c_2^{\mu})^2 + 2 \, c_3\, T_2 \, c_2^{\mu}
+\frac{1}{2} \, c_3^2 \, c_1^{\mu}
-\zeta_2 \, c_3^3 
-4\, c_2^{\mu} \, X_3 +2 \, c_3 \, X_2^{\mu}\right] \; 
\left[\frac{\ln^3(s_4/m_t^2)}{s_4}\right]_+
\nonumber \\ && 
{}+\left[\frac{3}{2}\, c_3 \,
c_2^{\mu} \, c_1^{\mu} 
+\frac{1}{2} \, (c_2^{\mu})^3
+\frac{3}{2} T_2 (c_2^{\mu})^2
-3\, \zeta_2 \, c_3^2 \,c_2
+\frac{5}{2} \, \zeta_3 \, c_3^3 \right.
\nonumber \\ && \hspace{10mm} \left.
{}+\frac{27}{2}\, \zeta_2 \, c_3 \, X_3 + 3\, c_2^{\mu} \, X_2^{\mu}
-\frac{3}{2} \, c_3 \, (X_1^{\mu^2}+X_1^{\zeta})
\right] \; \left[\frac{\ln^2(s_4/m_t^2)}{s_4}\right]_+   
\nonumber \\ && 
{}+\left[(c_2^{\mu})^2\, c_1^{\mu}-\zeta_2 \, c_3^2\, c_1^{\mu}
-\frac{5}{2} \, \zeta_2 \, c_3 \, \left((c_2^{\mu})^2+2\, T_2 \,
c_2^{\mu}\right)+5\, \zeta_3 \, c_3^2 \, c_2^{\mu}\right.
\nonumber \\ && \hspace{10mm} \left. 
{}+12\, \zeta_2 \, c_2^{\mu} \, X_3-5\, \zeta_2\, c_3\, X_2^{\mu}
-2\, c_2^{\mu}\, \left(X_1^{\mu^2}+X_1^{\zeta}\right)+c_3 X_0^{\zeta \, \mu}
\right] \; \left[\frac{\ln(s_4/m_t^2)}{s_4}\right]_+ 
\nonumber \\ && 
{}+\left[2 \zeta_3 c_3 (c_2^{\mu})^2
-\zeta_2 c_3 c_2^{\mu} c_1^{\mu}
-\frac{\zeta_2}{2} (c_2^{\mu})^2 \left(c_2^{\mu}+3 T_2 \right)
-3 \zeta_2 c_2^{\mu} X_2^{\mu}+\zeta_2 c_3 X_1^{\mu^2} \right.
\nonumber \\ && \hspace{10mm} \left. 
{}-c_1^{\mu} \, X_1^{\zeta \, \mu} 
+c_2^{\mu} \, X_0^{\zeta \, \mu}\right] \; \left[\frac{1}{s_4}\right]_+ 
\nonumber \\ && \left.
{}+\left[\frac{\zeta_3}{3} (c_2^{\mu})^3
-\frac{\zeta_2}{2}\left((c_2^{\mu})^2 c_1^{\mu}
-2 c_2^{\mu} X_1^{\mu^2}\right) \right]\delta(s_4)
\right\} \, ,
\label{NNNLO}
\eeqa
where, for the $t$ and $s$ channels, 
$X_3=\beta_0 \, c_3/12-C_F\beta_0/24$,
$X_2^{\mu}=(\beta_0/8)c_3\ln(\mu_R^2/m_t^2)$, 
$X_1^{\mu^2}=-(\beta_0/4)c_2^{\mu}
\ln(\mu_R^2/m_t^2)-C_F(\beta_0/4)\ln^2(\mu_F^2/m_t^2)$, 
$X_1^{\zeta}=(\beta_0/4)\, \zeta_2 \, c_3 
-C_F (\beta_0/8) \zeta_2$,
$X_1^{\zeta \,\mu}=-\zeta_2 c_3 c_2^{\mu}$, 
and 
$X_0^{\zeta \,\mu}=(\beta_0/8) \zeta_2 c_3 \ln(\mu_R^2/m_t^2)$.
The terminology is taken from Ref. \cite{NKNNNLO}. 
For associated $tW$ production the expressions for $X_2^{\mu}$,  
$X_1^{\zeta \,\mu}$, and $X_0^{\zeta \,\mu}$ are the same, but for the 
other variables we have $X_3^{tW}=\beta_0 \, c_3/12$, 
$X_1^{\mu^2\; tW}=-(\beta_0/4)c_2^{\mu}
\ln(\mu_R^2/m_t^2)-(C_F+C_A)(\beta_0/8)\ln^2(\mu_F^2/m_t^2)$, 
and $X_1^{\zeta\; tW}=(\beta_0/4)\, \zeta_2 \, c_3$. 
We note that only the leading and next-to-leading logarithms are complete.
Hence this is a NNNLO-NLL calculation.
Consistent with a NLL calculation \cite{NKtop,NKNNNLO} 
we have also kept all logarithms of the 
factorization and renormalization scales in the
$[\ln^3(s_4/m_t^2)/s_4]_+$ term, squares and cubes of logarithms
of the scales in the $[\ln^2(s_4/m_t^2)/s_4]_+$ term, and  
cubes of logarithms of the scales in the $[\ln(s_4/m_t^2)/s_4]_+$ term.
We have also kept  $\zeta_2$ and $\zeta_3$ terms
that arise in the calculation of the soft corrections 
when inverting from moments back to momentum space \cite{NKtop,NKNNNLO}.
This includes all $\zeta$ terms in the $[\ln^3(s_4/m_t^2)/s_4]_+$ 
and $[\ln^2(s_4/m_t^2)/s_4]_+$ terms, $\zeta$ terms multiplying logarithms 
of the scales in the $[\ln(s_4/m_t^2)/s_4]_+$ term, 
$\zeta$ terms multiplying squared and cubed logarithms of the scales in the 
$[1/s_4]_+$ term, and $\zeta$ terms multiplying cubed logarithms 
of the scales in the $\delta(s_4)$ term.

\mysection{Single top quark production via the $t$ channel at the Tevatron}

We now convolute the partonic cross sections in Section 3 with parton 
distribution functions (pdf) to obtain the hadronic cross section in 
$p{\bar p}$ collisions at the Tevatron.
Details on the kinematics are given in the Appendix. 
We use the MRST2004 NNLO pdf \cite{MRST2004} throughout. 
We also use standard values for the various electroweak parameters 
in the calculations (see the Appendix).

We begin with the $t$-channel. 
The dominant processes (with percentage contribution to the cross section) 
are $ub \rightarrow dt$ (65.7\%) and 
${\bar d} b \rightarrow {\bar u} t$ (21.4\%). Additional processes 
involving only quarks are $cb \rightarrow st$ (2.7\%) 
and the Cabibbo-suppressed $ub \rightarrow st$ (3.6\%), 
$cb \rightarrow dt$ (0.15\%) and $us \rightarrow dt$ (0.4\%); 
the contributions from even more suppressed processes 
($ub \rightarrow bt$, $cb \rightarrow bt$, $ud \rightarrow dt$, etc.) 
are negligible. Additional processes involving antiquarks and quarks 
are ${\bar s} b \rightarrow {\bar c}t$ (4.4\%) and the Cabibbo-suppressed 
${\bar d}b \rightarrow {\bar c}t$ (1.2\%), 
${\bar s}b \rightarrow {\bar u}t$ (0.2\%) and 
${\bar d} s \rightarrow {\bar u} t$ (0.14\%); the contributions from 
even more suppressed processes (${\bar s}s \rightarrow {\bar c}t$,
${\bar d}d \rightarrow {\bar u}t$, ${\bar s}d \rightarrow {\bar c}t$, etc.) 
are negligible.

\begin{table}[htb]
\begin{center}
\begin{tabular}{|c|c|c|c|c|} \hline
$t$ channel & LO & NLO approx & NNLO approx & NNNLO approx \\ \hline
$m_t=170$  & 1.131 & 1.150  & 1.177  & 1.193 \\ \hline
$m_t=172$  & 1.091 & 1.113  & 1.139  & 1.155 \\ \hline
$m_t=175$ & 1.035 & 1.060 & 1.085 & 1.100 \\ \hline
\end{tabular}
\caption[]{The leading-order and approximate higher-order 
cross sections for top quark production in the $t$ channel in pb 
for $p \overline p$ collisions with $\sqrt{S} = 1.96$ TeV and 
$m_t=170$, 172, and 175 GeV. We use the MRST2004 NNLO pdf and we set 
$\mu_F=\mu_R=m_t$.}
\end{center}
\end{table}

In Table 1 we give results for the leading-order (LO) cross section and for  
the approximate NLO, NNLO, and NNNLO cross sections that include the threshold 
corrections at NLL accuracy at each order (i.e. NLO-NLL, NNLO-NLL, NNNLO-NLL).
We set $\mu_F=\mu_R=m_t$ and show results for three different top-quark 
mass values.
We use the same pdf set for all results because we are interested in the size 
of the terms at each order in the perturbative calculation with all other 
things held constant.  
Note that the soft-gluon corrections are relatively small for this channel. 
This is also true for the exact NLO corrections \cite{bwhl}. 
In fact, the approximate 
NLO cross section is less than 2\% larger than the exact NLO result.

The most recent value for the top quark mass from the Tevatron is 
$m_t=171.4 \pm 2.1$ GeV \cite{topmass}, 
so it is interesting to calculate the cross section 
for this specific mass value.  
The NNNLO approximate cross section at 171.4 GeV is 1.17 pb. 
If we match this to the exact NLO cross section, then the matched 
cross section (i.e. exact NLO plus NNLO and NNNLO threshold corrections) is  
\beq
\sigma^{t-{\rm channel}}(m_t=171.4 \pm 2.1 \,{\rm GeV})=1.15^{+0.02}_{-0.01} 
\pm 0.04 \, \pm 0.06 \; {\rm pb}\, ,
\eeq
where the first uncertainty is due to the scale dependence, 
the second is due to the mass, and the third is due to the pdf. 
Adding these uncertainties in quadrature we find that 
$\sigma^{t-{\rm channel}}(m_t=171.4 \pm 2.1 \,{\rm GeV})
=1.15 \pm 0.07$ pb.
A few more remarks are in order regarding the uncertainties. 
The scale uncertainty 
results by varying the scale between $m_t/2$ and $2m_t$. Although this is a 
standard procedure, theoretically it is not unambiguous. The mass uncertainty 
in the cross section is found  by using the experimentally determined 
uncertainty in the mass of the top quark ($\pm 2.1$ GeV). Regarding the pdf 
uncertainty, since the MRST2004 densities do not come with errors, we use 
instead the pdf uncertainty from the MRST2001E NLO pdf \cite{MRST2001E} 
(the two sets give very similar values for the cross section and the latter 
set also provides pdf uncertainties).

It is also important to provide the matched cross section for $m_t=175$ GeV, 
a top quark mass value that has been used widely in cross section calculations.
As shown in Table 1, the NNNLO approximate cross section at 175 GeV is 1.100 
pb. After matching to the exact NLO cross section we find 
$\sigma^{t-{\rm channel}}(m_t=175 \,{\rm GeV})=1.08^{+0.02}_{-0.01}\pm 0.06$ 
pb, where the first uncertainty is due to the scale dependence and
the second is due to the pdf, and no uncertainty is considered for the mass.
Adding these uncertainties in quadrature we find 
 $\sigma^{t-{\rm channel}}(m_t=175 \,{\rm GeV})=1.08 \pm 0.06$ pb.

\begin{figure}
\begin{center}
\includegraphics[width=11cm]{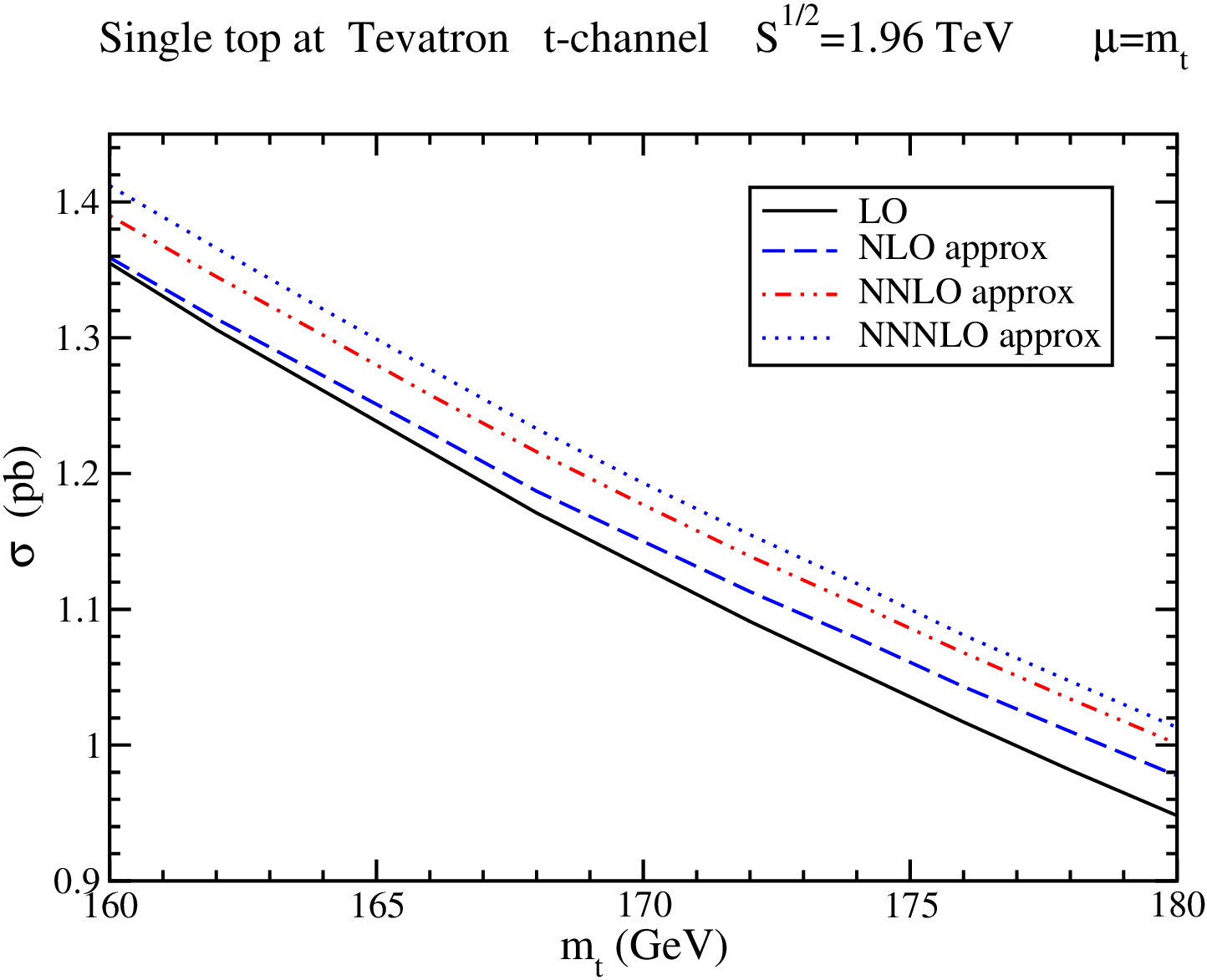}
\caption{The cross section for single top quark production at the Tevatron 
in the $t$ channel. Here $\mu=\mu_F=\mu_R=m_t$.}
\label{tchtevmtplot}
\end{center}
\end{figure}

In Fig. \ref{tchtevmtplot} we plot the cross section 
for single top quark production at the Tevatron with $\sqrt{S}=1.96$ TeV 
in the $t$ channel using the MRST2004 NNLO parton densities and setting 
both the factorization and renormalization scales to a common scale $\mu=m_t$. 
We plot the LO cross section 
and the approximate NLO, NNLO, and NNNLO cross sections at NLL accuracy.

\begin{figure}
\begin{center}
\includegraphics[width=11cm]{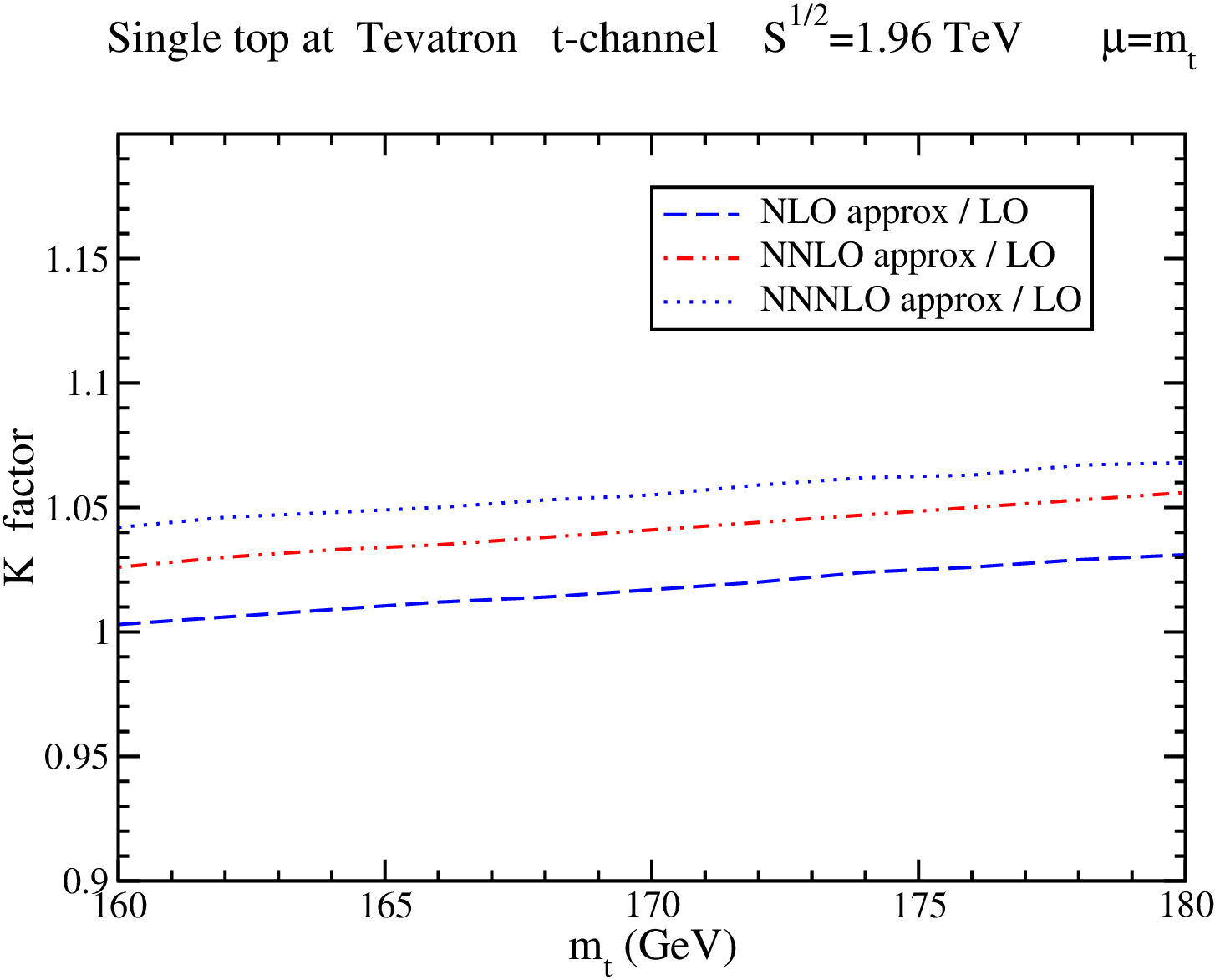}
\caption{The $K$ factors for single top quark production at the Tevatron 
in the $t$ channel. Here $\mu=\mu_F=\mu_R=m_t$.}
\label{Ktchtevmtplot}
\end{center}
\end{figure}

The relative contribution from higher orders is shown in 
Fig. \ref{Ktchtevmtplot} where the $K$ factors, defined as ratios of 
the higher-order cross sections to LO, are shown. The $K$ factors 
in the $t$ channel are small, 
showing that the corrections do not provide a big enhancement to 
the cross section.

\begin{figure}
\begin{center}
\includegraphics[width=11cm]{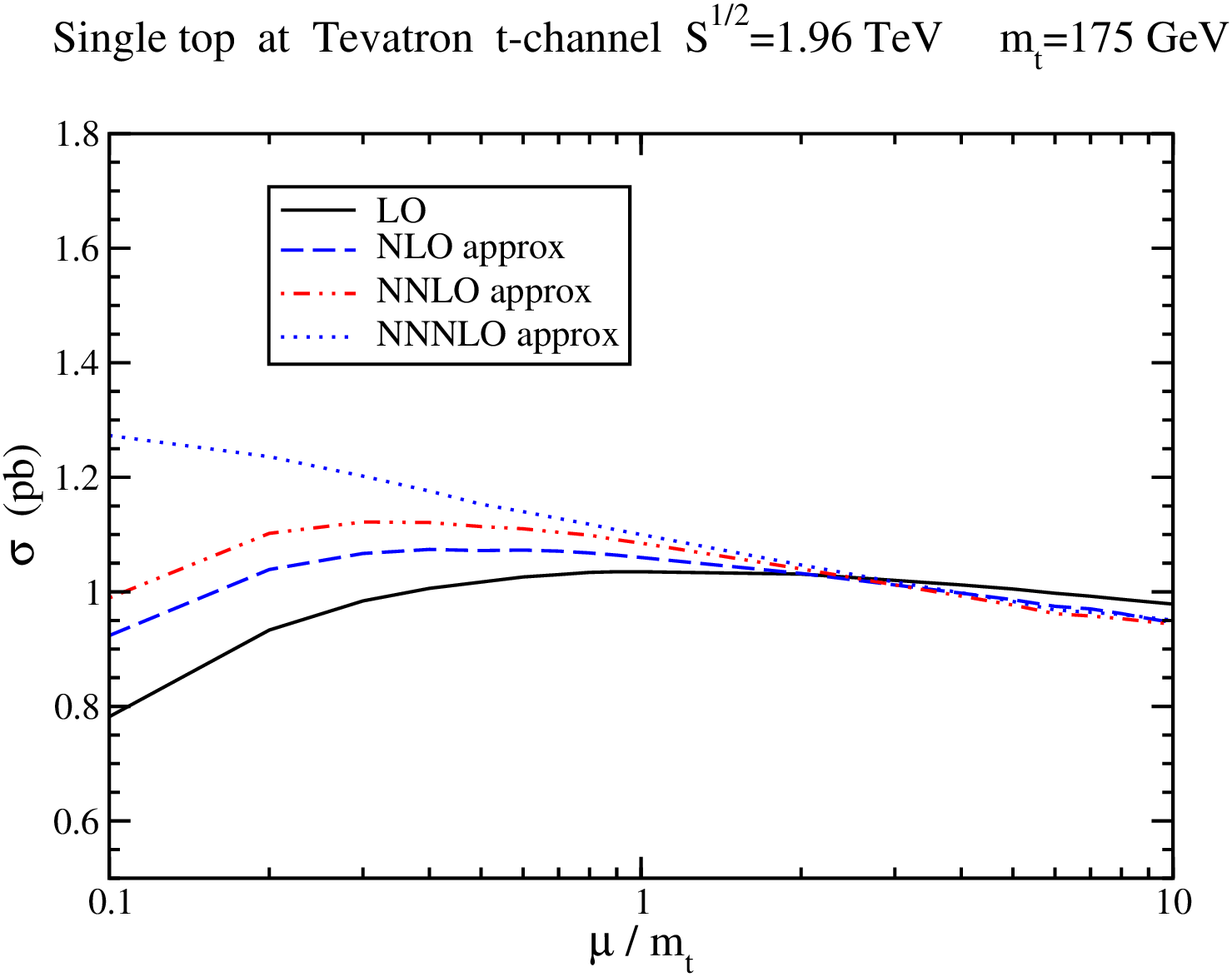}
\caption{Scale dependence of the single top quark cross section at the 
Tevatron in the $t$ channel. Here $\mu=\mu_F=\mu_R$.}
\label{tchtevmu175plotdelta}
\end{center}
\end{figure}

In Fig. \ref{tchtevmu175plotdelta} we plot the scale dependence 
of the cross section with $m_t=175$ GeV. 
We set the factorization scale equal to the 
renormalization scale and vary this common scale $\mu$ over two orders 
of magnitude. We show results for the LO and higher-order cross sections.

\begin{figure}
\begin{center}
\includegraphics[width=11cm]{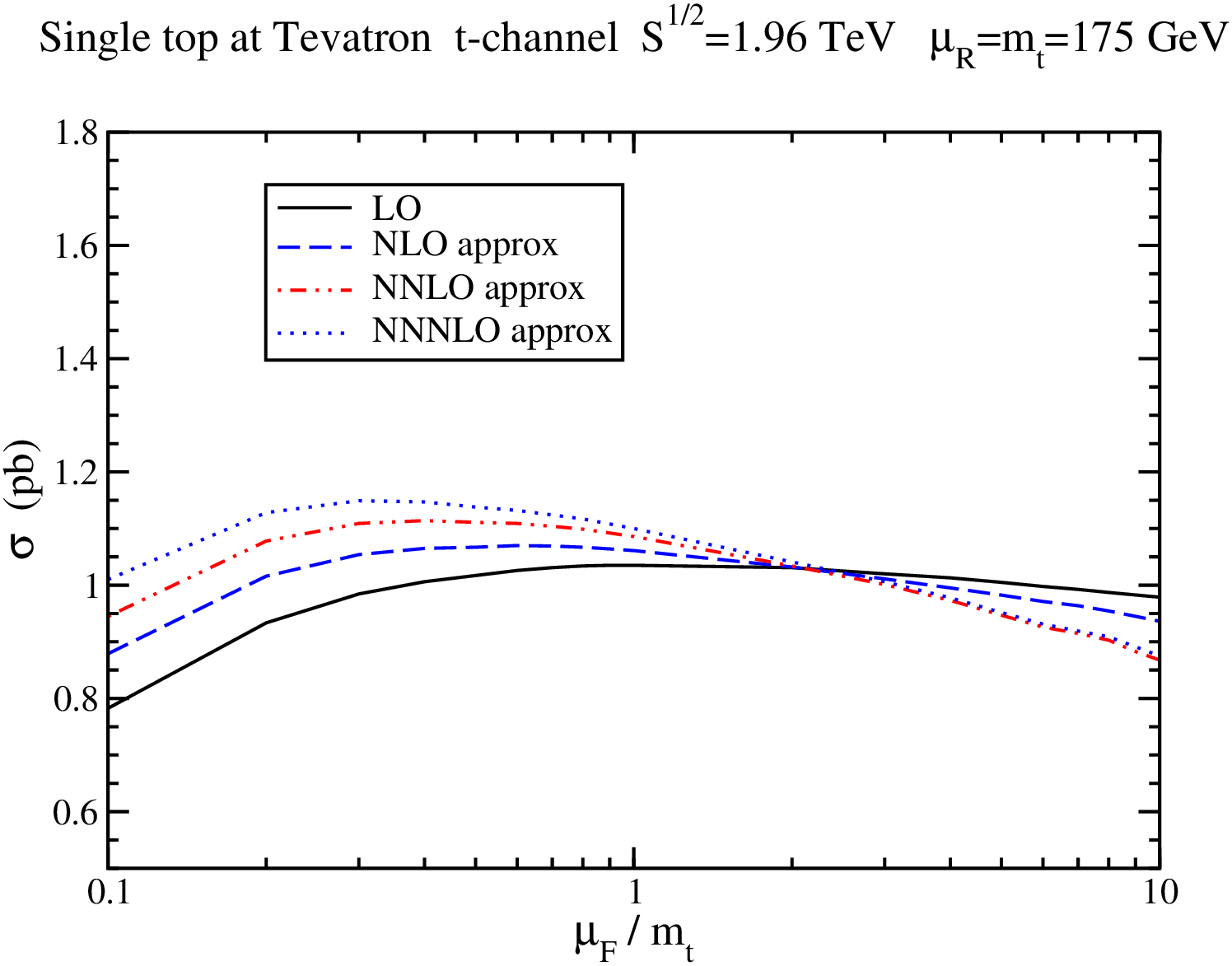}
\caption{Factorization scale dependence of the single top quark cross section 
at the Tevatron in the $t$ channel. Here the renormalization scale is 
held fixed, $\mu_R=m_t=175$ GeV.}
\label{tchtevmuF175plotdelta}
\end{center}
\end{figure}

In general the factorization and renormalization scales are independent 
and it is interesting to see the dependence of the cross section separately
on each scale.
In Fig. \ref{tchtevmuF175plotdelta} we plot the factorization scale, 
$\mu_F$, dependence of the cross section,  while setting the 
renormalization scale $\mu_R=m_t$. 
We vary $\mu_F$ over two orders of magnitude.
At LO the result is exactly the same as in  Fig. \ref{tchtevmu175plotdelta} 
because at LO there is no $\mu_R$ dependence. However at higher orders 
the difference is noticable.

\begin{figure}
\begin{center}
\includegraphics[width=11cm]{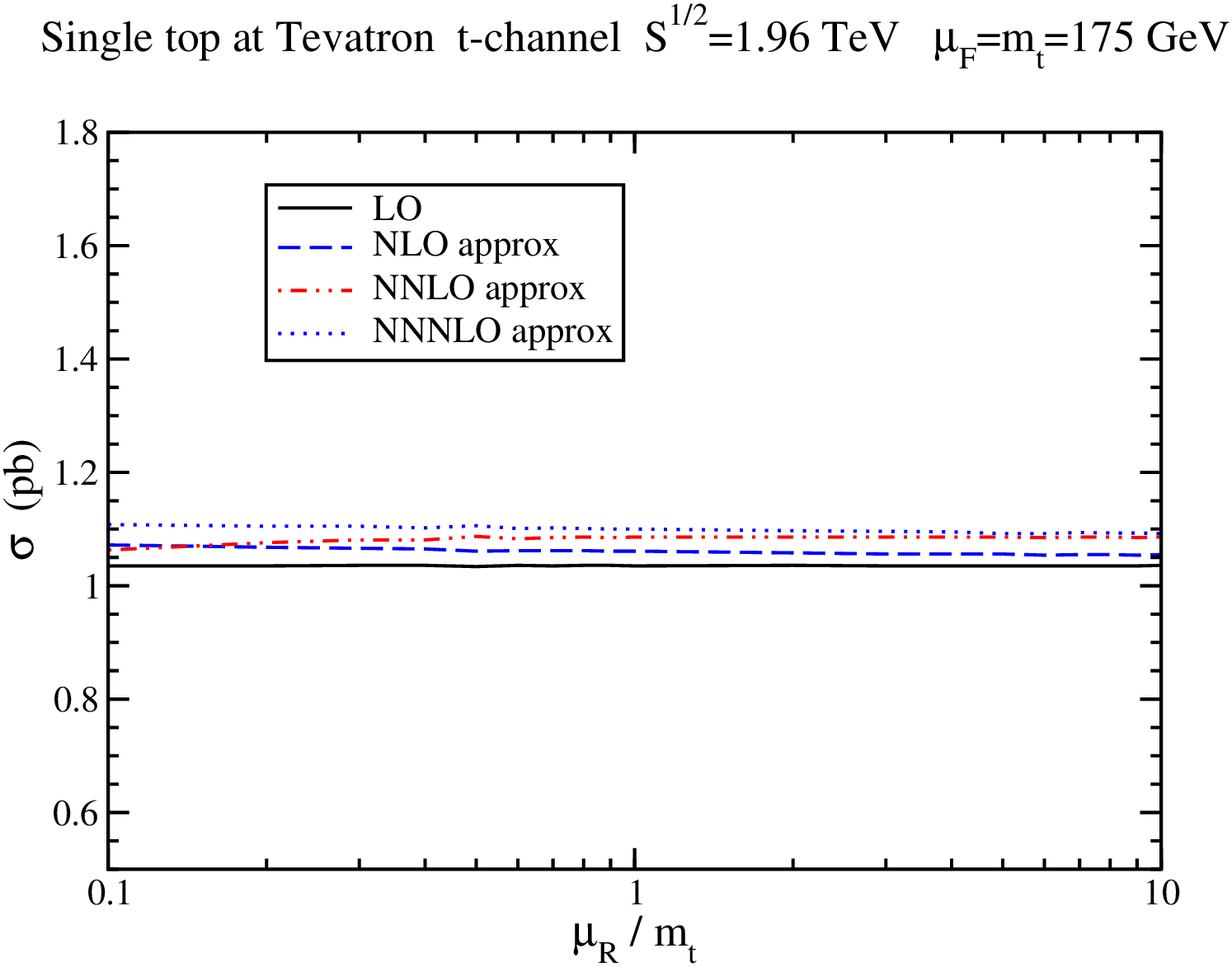}
\caption{Renormalization scale dependence of the single top quark cross 
section at the Tevatron in the $t$ channel. Here the factorization scale is 
held fixed, $\mu_F=m_t=175$ GeV.}
\label{tchtevmuR175plot}
\end{center}
\end{figure}

In Fig. \ref{tchtevmuR175plot} we plot the renormalization scale, 
$\mu_R$, dependence of the cross section,  while setting the 
factorization scale $\mu_F=m_t$. 
We vary $\mu_R$ over two orders of magnitude.
At LO the result is exactly flat because, as noted above, at LO there is 
no $\mu_R$ dependence. At higher orders there is a $\mu_R$ dependence, 
though it is fairly flat compared 
to the $\mu_F$ dependence in Fig. \ref{tchtevmuF175plotdelta}.

\mysection{Single top quark production via the $s$ channel at the Tevatron}

We continue with the $s$ channel.
The dominant process (with percentage contribution to the cross section) 
is $u {\bar d} \rightarrow {\bar b} t$ (97.4\%).
Additional processes are $c {\bar s} \rightarrow {\bar b} t$ (1.1\%) 
and the Cabibbo-suppressed $u {\bar s} \rightarrow {\bar b} t$ (1.1\%), 
$c{\bar d} \rightarrow {\bar b}t$ (0.26\%), and  
$u{\bar d} \rightarrow {\bar s}t$ (0.17\%); 
the contributions from even more suppressed processes 
($u {\bar d} \rightarrow {\bar d}t$, $c {\bar s} \rightarrow {\bar d} t$, etc.)
are negligible.

\begin{table}[htb]
\begin{center}
\begin{tabular}{|c|c|c|c|c|} \hline
$s$ channel & LO & NLO approx & NNLO approx & NNNLO approx \\ \hline
$m_t=170$  & 0.353 & 0.510  & 0.555  & 0.576 \\ \hline
$m_t=172$  & 0.335 & 0.484  & 0.528  & 0.547 \\ \hline
$m_t=175$ & 0.310 & 0.448 & 0.488 & 0.506 \\ \hline
\end{tabular}
\caption[]{The leading-order and approximate higher-order 
cross sections for top quark production in the $s$ channel in pb 
for $p \overline p$ collisions with $\sqrt{S} = 1.96$ TeV and 
$m_t=170$, 172, and 175 GeV. We use the MRST2004 NNLO pdf and we set 
$\mu_F=\mu_R=m_t$.}
\end{center}
\end{table}

In Table 2 we give results for the LO cross section and for 
the approximate NLO, NNLO, and NNNLO cross sections with the threshold 
corrections at NLL accuracy.
The soft-gluon corrections are relatively large for this channel, in 
stark contrast with the results we found in the $t$ channel. 
This is also true for the exact NLO cross section \cite{bwhl}. 
That the behavior of the two channels is quite different should not be 
surprising: the kinematics and the color flows are quite different.
The $t$ channel resembles deep inelastic scattering while the $s$ channel 
resembles the Drell-Yan process.
As we will see, there are differences between the channels not only 
in the $K$ factors but also in the scale dependence.
We also note that the approximate NLO cross section in the $s$ channel 
is only 3\% larger than the exact NLO result, showing that 
the threshold corrections are dominant and provide the bulk of the QCD 
corrections and, thus, that the threshold approximation works very well.

Again, it is interesting to provide the cross section for the specific value 
of the new mass from the Tevatron, $m_t=171.4 \pm 2.1$ GeV.  
The NNNLO approximate cross section at 171.4 GeV is 0.555 pb. 
If we match this to the exact NLO cross section, then the matched 
cross section (i.e. exact NLO plus NNLO and NNNLO threshold corrections) is
\beq
\sigma^{s-{\rm channel}}(m_t=171.4 \pm 2.1 \, {\rm GeV})=0.54 \pm 0.02 
\pm 0.03 \, \pm 0.01 \; {\rm pb}\, ,
\eeq
where the first uncertainty is due to scale variation between $m_t/2$ and 
$2m_t$, the second is due to the mass ($\pm 2.1$ GeV), and the third is the 
pdf uncertainty, as discussed in the previous section.
Adding these uncertainties in quadrature we find that 
$\sigma^{s-{\rm channel}}(m_t=171.4 \pm 2.1 \, {\rm GeV})=0.54 \pm 0.04$ pb.

As before, it is also important to provide the matched cross section for 
$m_t=175$ GeV.
As shown in Table 2, the NNNLO approximate cross section at 175 GeV is 0.506 
pb. After matching to the exact NLO cross section we find 
$\sigma^{s-{\rm channel}}(m_t=175 \,{\rm GeV})=0.49 \pm 0.02 \pm 0.01$ pb, 
where the first uncertainty is due to the scale dependence and
the second is due to the pdf, and no uncertainty is considered for the mass.
Adding these uncertainties in quadrature we find 
 $\sigma^{s-{\rm channel}}(m_t=175 \,{\rm GeV})=0.49 \pm 0.02$ pb.

\begin{figure}
\begin{center}
\includegraphics[width=11cm]{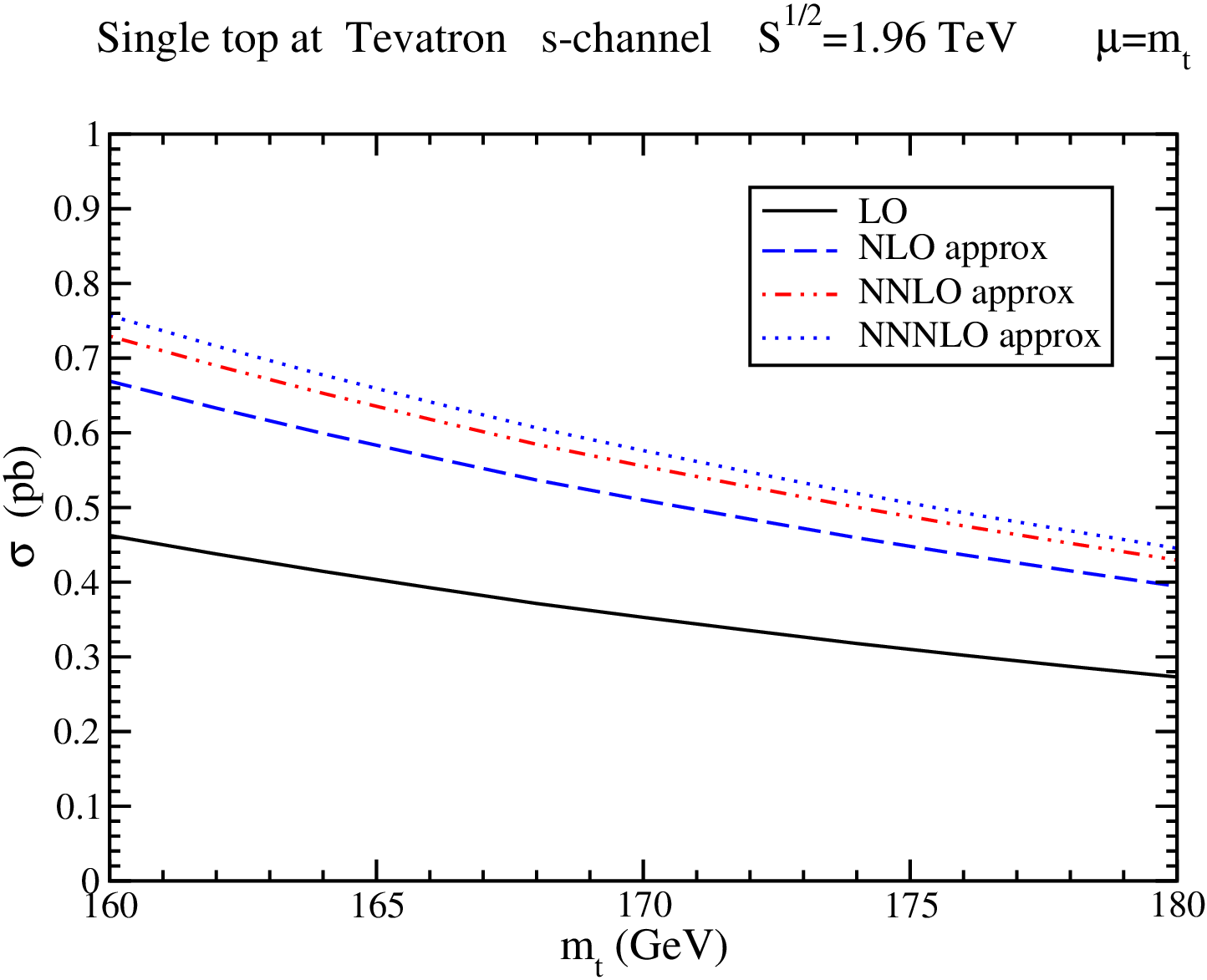}
\caption{The cross section for single top quark production at the Tevatron 
in the $s$ channel. Here $\mu=\mu_F=\mu_R=m_t$.}
\label{schtevmtplot}
\end{center}
\end{figure}

In Fig. \ref{schtevmtplot} we plot the cross section 
for single top quark production at the Tevatron with $\sqrt{S}=1.96$ TeV 
in the $s$ channel using the MRST2004 parton densities and setting 
both the factorization and renormalization scales to $\mu=m_t$. 
We plot the LO cross section 
and the approximate NLO, NNLO, and NNNLO cross sections at NLL accuracy.

\begin{figure}
\begin{center}
\includegraphics[width=11cm]{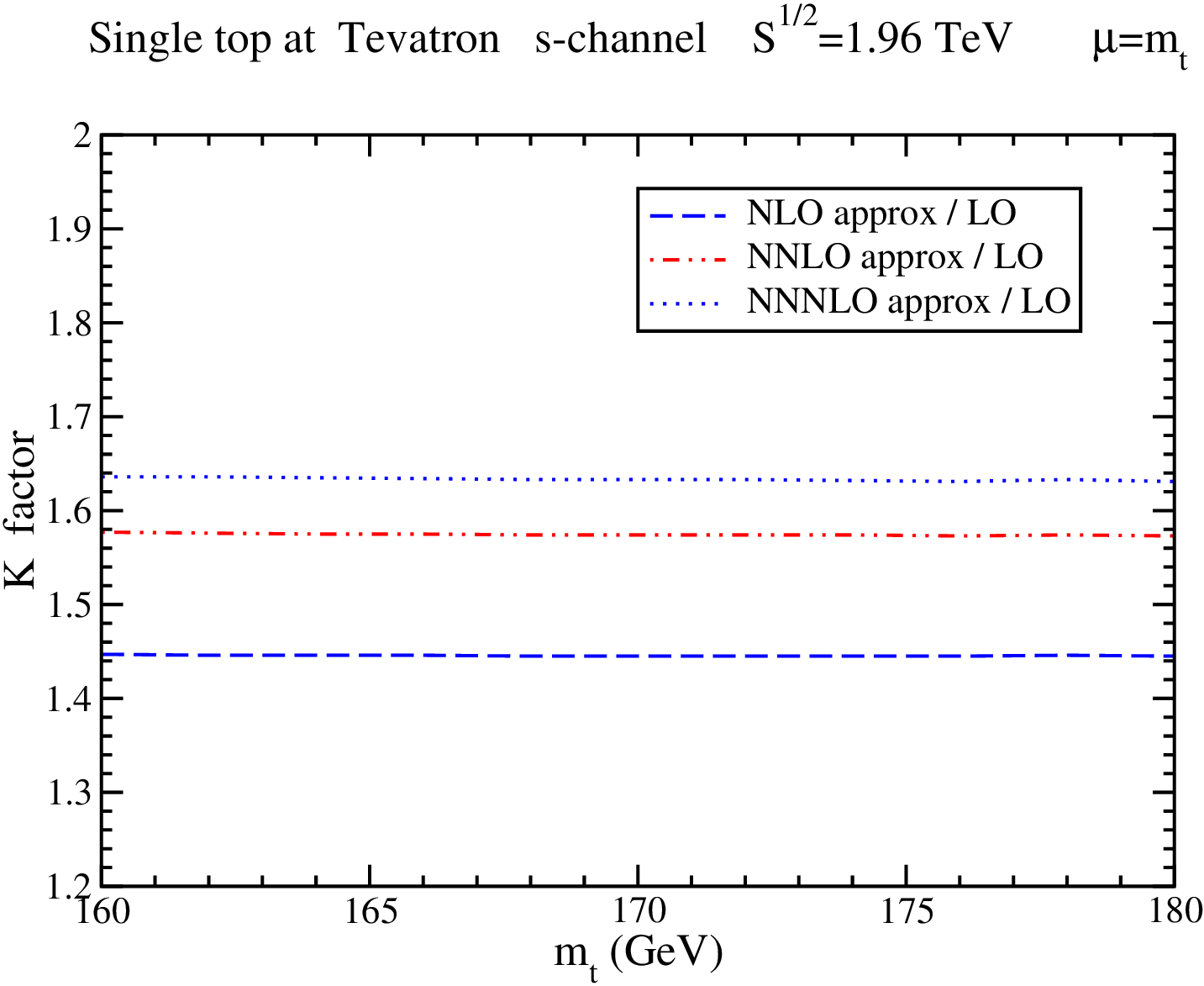}
\caption{The $K$ factors for single top quark production at the Tevatron 
in the $s$ channel. Here $\mu=\mu_F=\mu_R=m_t$.}
\label{Kschtevmtplot}
\end{center}
\end{figure}

The $K$ factors are shown in Fig. \ref{Kschtevmtplot}. 
They are quite large, thus showing that the corrections provide a 
big enhancement to the cross section.
This behavior is to be contrasted with the $t$ channel where the 
corrections are small. We also note that the $K$-factors in the 
$s$ channel are fairly 
constant over the top-quark mass range shown. As seen in the plot, 
the NLO corrections provide a 45\% increase of the LO cross section, the 
NNLO corrections provide an additional 12\%, and the NNNLO corrections 
a further 6\%.

\begin{figure}
\begin{center}
\includegraphics[width=11cm]{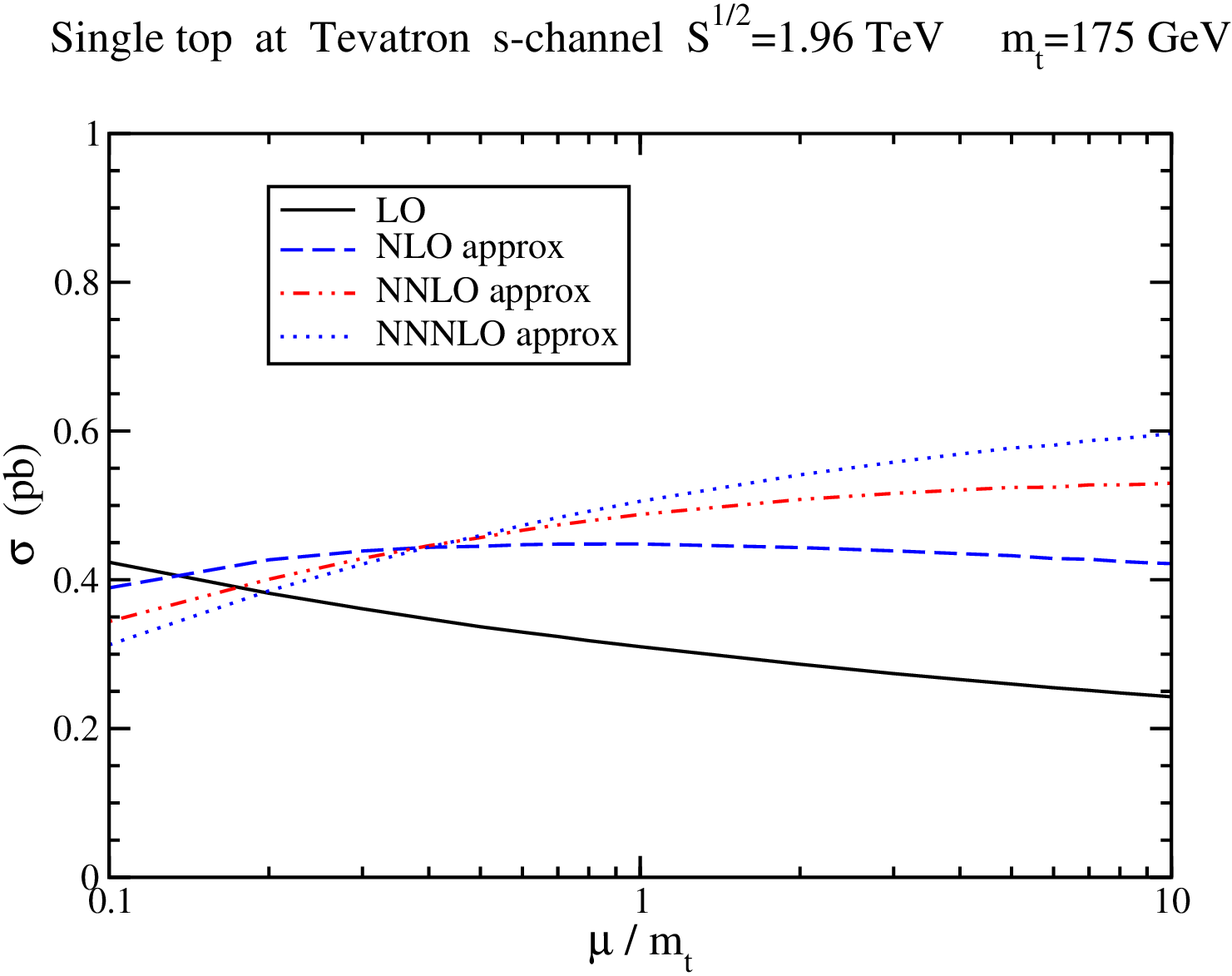}
\caption{Scale dependence of the single top quark cross section at the 
Tevatron in the $s$ channel. Here $\mu=\mu_F=\mu_R$.}
\label{schtevmu175plotdelta}
\end{center}
\end{figure}

In Fig. \ref{schtevmu175plotdelta} we plot the scale dependence 
of the cross section with $m_t=175$ GeV. 
We set the factorization scale equal to the 
renormalization scale and vary this common scale $\mu$ over two orders 
of magnitude. We show results for the LO and higher-order cross sections.

\begin{figure}
\begin{center}
\includegraphics[width=11cm]{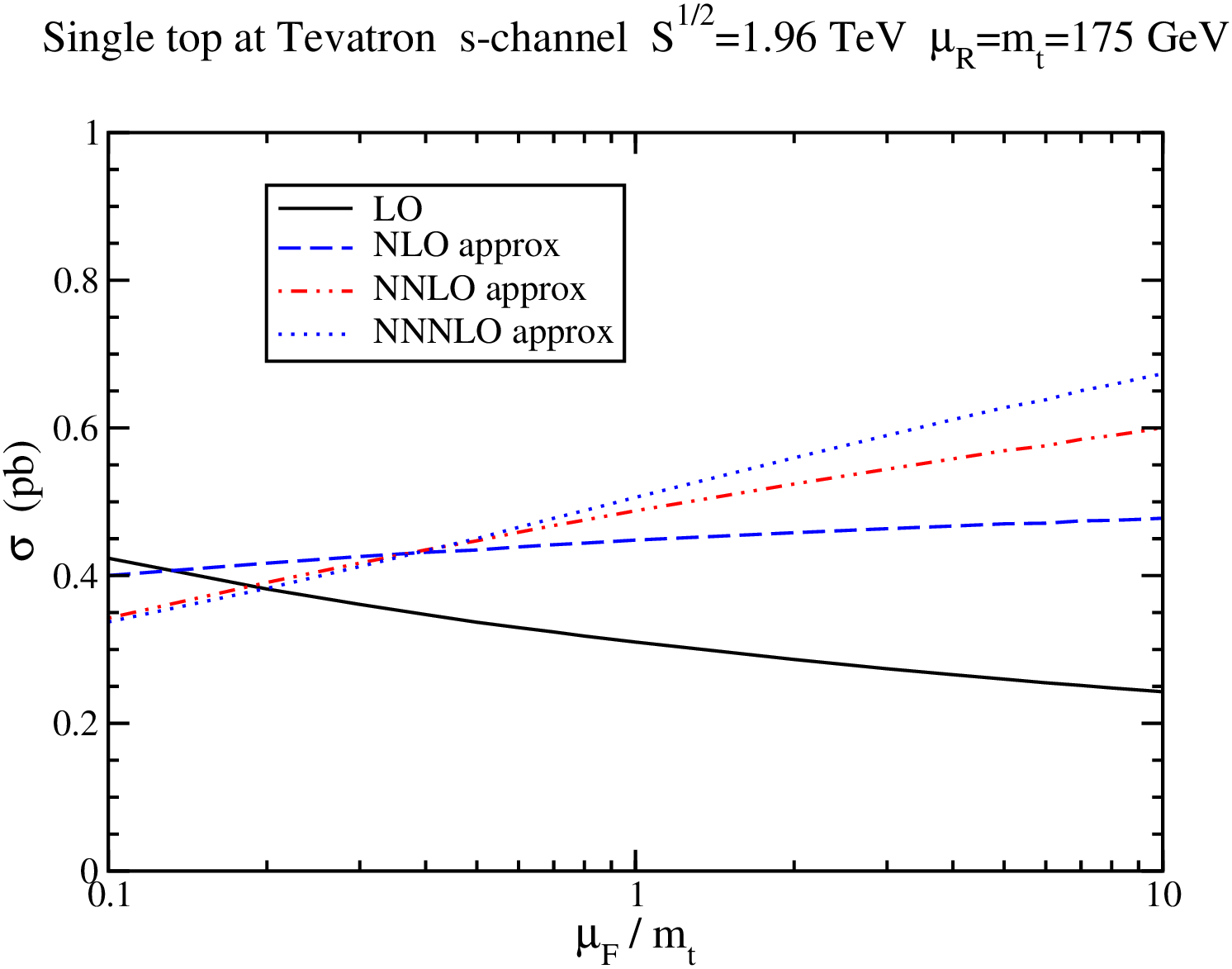}
\caption{Factorization scale dependence of the single top quark cross section 
at the Tevatron in the $s$ channel. Here the renormalization scale is 
held fixed, $\mu_R=m_t=175$ GeV.}
\label{schtevmuF175plotdelta}
\end{center}
\end{figure}

Again, we note that the factorization and renormalization scales are 
theoretically independent, and it is interesting to see the dependence 
of the cross section separately on each scale.
In Fig. \ref{schtevmuF175plotdelta} we plot the factorization scale, 
$\mu_F$, dependence of the cross section,  while setting the 
renormalization scale $\mu_R=m_t$. 
We vary $\mu_F$ over two orders of magnitude.
At LO the result is exactly the same as in  Fig. \ref{schtevmu175plotdelta} 
because of the lack of $\mu_R$ dependence. 

\begin{figure}
\begin{center}
\includegraphics[width=11cm]{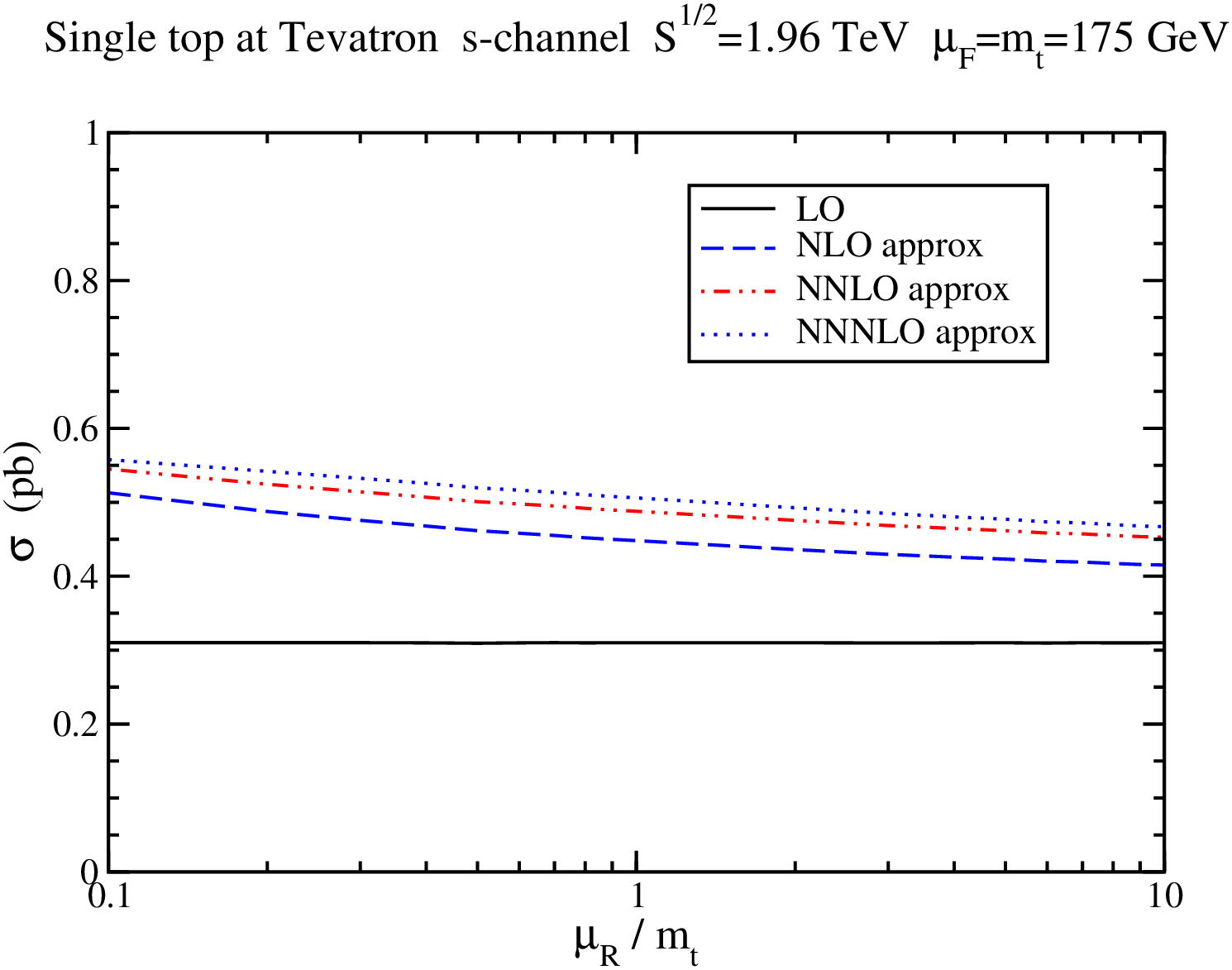}
\caption{Renormalization scale dependence of the single top quark cross 
section at the Tevatron in the $s$ channel. Here the factorization scale is 
held fixed, $\mu_F=m_t=175$ GeV.}
\label{schtevmuR175plot}
\end{center}
\end{figure}

In Fig. \ref{schtevmuR175plot} we plot the renormalization scale, 
$\mu_R$, dependence of the cross section,  while setting the 
factorization scale $\mu_F=m_t$. 
We vary $\mu_R$ over two orders of magnitude.
At LO the result is exactly flat, as noted before. 
At higher orders there is a $\mu_R$ dependence and it is less flat 
than for the $t$ channel.

\mysection{Associated $tW$ production at the Tevatron}

Associated $tW$ production proceeds via $bg \rightarrow tW^-$ 
(Cabibbo-suppressed contributions from 
$sg\rightarrow tW^-$ and $dg\rightarrow tW^-$ are negligible). 
The kinematics and color flow for this process are rather different 
from both the $t$ and $s$ channels since we have a real $W$ boson 
in the final state and a quark-gluon vertex at lowest order. 

The LO cross section is 0.070 pb for $m_t=175$ GeV, which is a rather 
small cross section and thus makes this channel relatively unimportant 
at the Tevatron even after we have added the threshold corrections.

\begin{table}[htb]
\begin{center}
\begin{tabular}{|c|c|c|c|c|} \hline
$tW$ production & LO & NLO approx & NNLO approx & NNNLO approx \\ \hline
$m_t=170$  & 0.080 & 0.122 & 0.140 & 0.146 \\ \hline
$m_t=172$  & 0.076 & 0.116 & 0.133 & 0.139 \\ \hline
$m_t=175$  & 0.070 & 0.107 & 0.122 & 0.127 \\ \hline
\end{tabular}
\caption[]{The leading-order and approximate higher-order 
cross sections for associated $tW$ production in pb 
for $p \overline p$ collisions with $\sqrt{S} = 1.96$ TeV and 
$m_t=170$, 172, and 175 GeV. We use the MRST2004 NNLO pdf and we set  
$\mu_F=\mu_R=m_t$.}
\end{center}
\end{table}

In Table 3 we give results for the LO and 
the approximate NLO, NNLO, and NNNLO cross sections that include the threshold 
corrections at NLL accuracy at each order.
The soft-gluon corrections are relatively large for this channel, even  
more than in the $s$ channel. 

Again, it is interesting to give a result for the specific value of 
the new mass from the Tevatron, $m_t=171.4 \pm 2.1$ GeV.  
The NNNLO approximate cross section at 171.4 GeV is 
\beq
\sigma^{tW}(m_t=171.4 \pm 2.1 \, {\rm GeV})=0.14 \pm 0.02 ^{+0.01}_{-0.02}\, 
\pm 0.02 \; {\rm pb}\, ,
\eeq
where the first uncertainty is due to scale variation between $m_t/2$ and 
$2m_t$, the second is due to the mass ($\pm 2.1$ GeV), and the third is the 
pdf uncertainty, as discussed previously.
Adding these uncertainties in quadrature we find that 
$\sigma^{tW}(m_t=171.4 \pm 2.1 \, {\rm GeV})=0.14 \pm 0.03$ pb.
We note that since an exact NLO result is not available we do not provide a 
matched cross section here.

As before, it is also important to provide the cross section for 
$m_t=175$ GeV.
As shown in Table 3, the NNNLO approximate cross section at 175 GeV is 0.127 
pb. Including uncertainties we have 
$\sigma^{tW}(m_t=175 \,{\rm GeV})=0.13 \pm 0.02 \pm 0.02$ pb, 
where the first uncertainty is due to the scale dependence and
the second is due to the pdf, and no uncertainty is considered for the mass.
Adding these uncertainties in quadrature we find 
 $\sigma^{tW}(m_t=175 \,{\rm GeV})=0.13 \pm 0.03$ pb.

\begin{figure}
\begin{center}
\includegraphics[width=11cm]{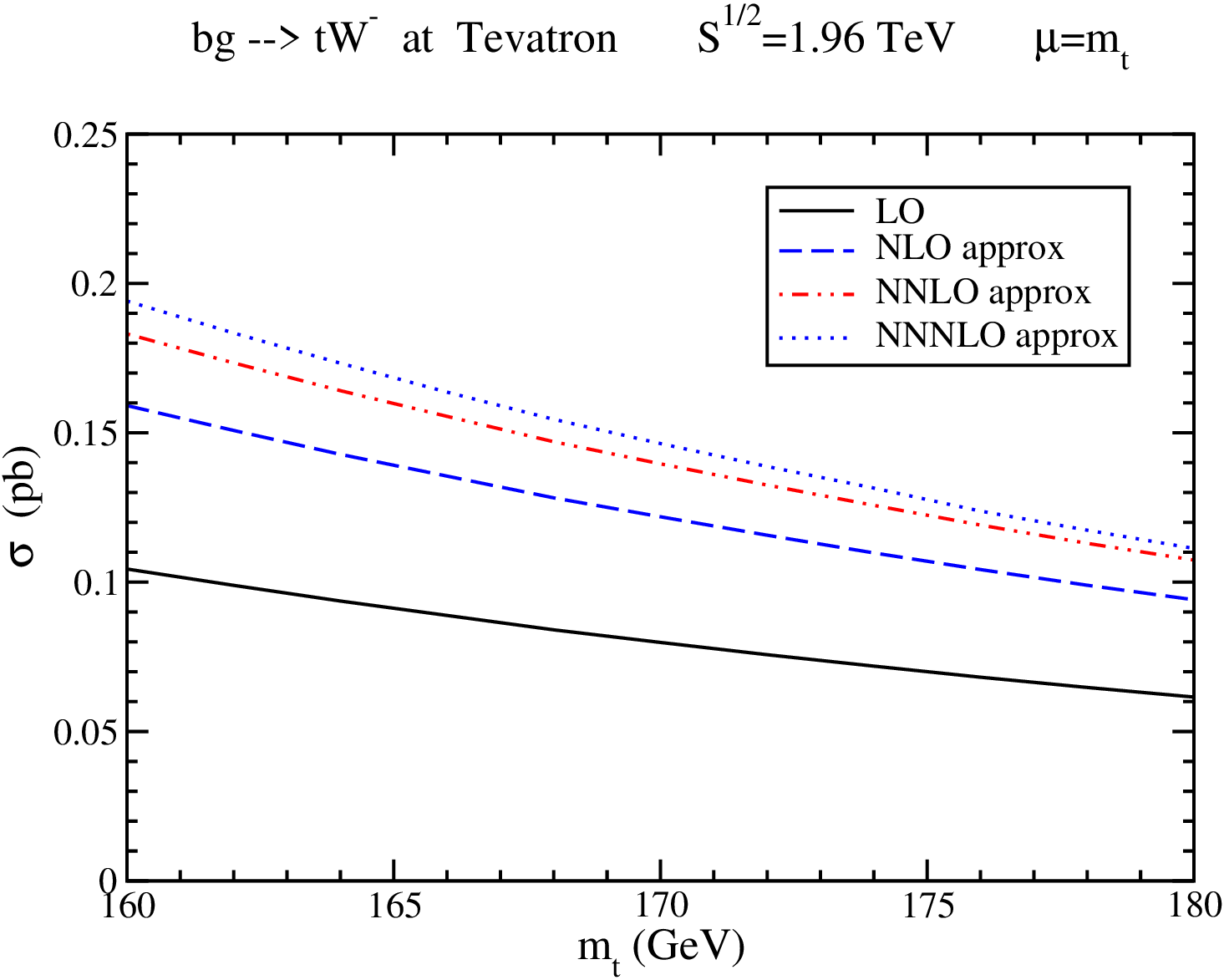}
\caption{The cross section for associated $tW$ production at the Tevatron. 
Here $\mu=\mu_F=\mu_R=m_t$.}
\label{bgtevmtplot}
\end{center}
\end{figure}

In Fig. \ref{bgtevmtplot} we plot the cross section 
for single top production at the Tevatron with $\sqrt{S}=1.96$ TeV 
via associated $tW$ production using the MRST2004 parton densities and setting 
both the factorization and renormalization scales to $\mu=m_t$. 
We plot the LO cross section 
and the approximate NLO, NNLO, and NNNLO cross sections at NLL accuracy.

\begin{figure}
\begin{center}
\includegraphics[width=11cm]{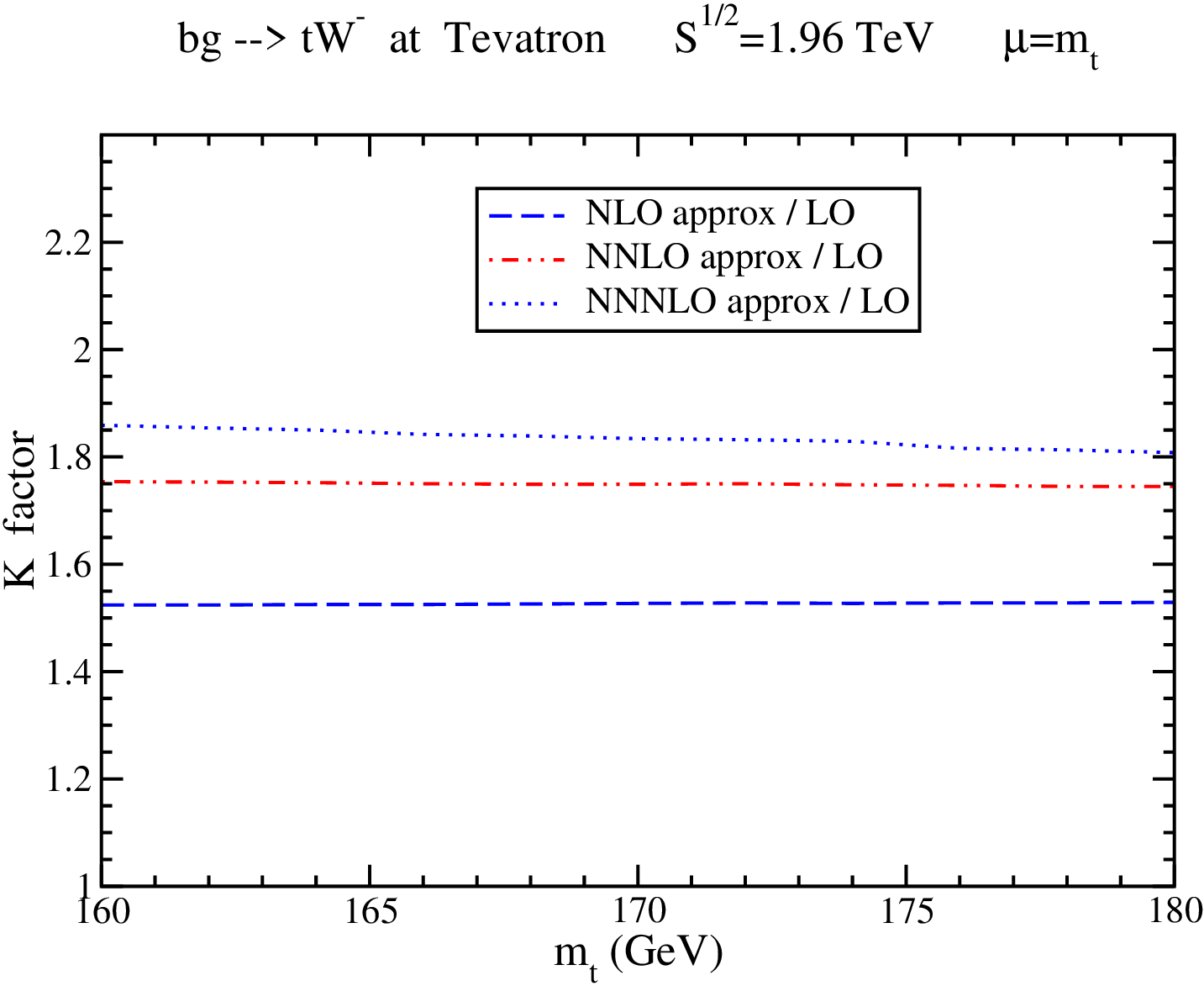}
\caption{The $K$ factors for associated $tW$ production at the Tevatron. 
Here $\mu=\mu_F=\mu_R=m_t$.}
\label{Kbgtevmtplot}
\end{center}
\end{figure}

The relative contribution from higher orders is shown in 
Fig. \ref{Kbgtevmtplot}. 
The $K$ factors are quite large, showing that the soft-gluon corrections 
provide a big enhancement to the cross section, and fairly 
constant over the top-quark mass range shown. As seen from the plot, 
the NLO corrections provide a 53\% increase of the LO cross section, the 
NNLO corrections provide an additional 22\%, and the NNNLO corrections 
a further 8\%.
However, the overall cross section still remains rather small and we will 
not study it further here.

\mysection{Conclusion}

We have studied single top quark production at the Tevatron via all 
Standard Model partonic processes. We have presented the resummation 
of the threshold soft-gluon corrections and their expansion to 
provide higher-order approximate cross sections through NNNLO.

The results differ a lot among channels. In the $t$ channel the 
soft-gluon corrections are small and hence do not greatly change the 
LO result. Including soft gluon corrections through NNNLO at NLL 
accuracy and matching with the exact NLO cross section, our best estimate 
of the cross section is  
$\sigma^{t-{\rm channel}}(m_t=171.4 \pm 2.1\,{\rm GeV})
=1.15 \pm 0.07$ pb, 
where the uncertainty indicated includes the scale dependence, 
the uncertainty in the top quark mass, and the pdf uncertainty.

In the $s$ channel, however, the corrections are significant. The NLO 
approximate cross section is an excellent approximation to the full 
NLO result, thus showing that the threshold approximation works very 
well, while the higher-order contributions provide further enhancements.
Our best estimate for the cross section in this channel is 
$\sigma^{s-{\rm channel}}(m_t=171.4 \pm 2.1 \, {\rm GeV})=0.54 \pm 0.04$ pb.

The renormalization and factorization scale dependence of the cross 
section for both $t$ and $s$ channels  were displayed both separately and 
also setting the scales equal to each other. 

The threshold corrections for associated $tW$ production are large as a 
percentage of the LO cross section (large $K$ factors), however even with 
those the total cross section is still rather small for this channel.
We find $\sigma^{tW}(m_t=171.4 \pm 2.1 \, {\rm GeV})=0.14 \pm 0.03$ pb.

Finally, we note that the corresponding cross sections for single antitop 
quark production at the Tevatron are identical to those for 
single top quark production as presented in this paper.

\mysection*{Acknowledgements}

This work has been supported by the National Science Foundation under 
grant PHY 0555372.

\setcounter{equation}{0}
\renewcommand{\theequation}{A.\arabic{equation}}

\newpage

\mysection*{Appendix}

Here we provide some details on the kinematics of the calculation.
We consider the process $p_1+p_2 \rightarrow p_3+p_4$, where the particles 
have been represented by their momenta $p_i$.
If $p_1$ and $p_2$ refer to incoming partons (quarks or gluons) then $m_1=0$ 
and $m_2=0$. The partonic kinematical invariants are 
$s=(p_1+p_2)^2$, $t=(p_1-p_3)^2$, $u=(p_2-p_3)^2$, 
$s_4=s+t+u-m_3^2-m_4^2$.
The hadronic kinematical invariants are 
$S=(p_{h1}+p_{h2})^2$, $T=(p_{h1}-p_3)^2$, $U=(p_{h2}-p_3)^2$, 
$S_4=S+T+U-m_3^2-m_4^2$, where $h_1$ and $h_2$ are the 
hadrons (protons and antiprotons at the Tevatron) corresponding to partons 
$p_1$ and $p_2$; thus, $p_1=x_1 p_{h1}$ and $p_2=x_2 p_{h2}$.
The relations between partonic and hadronic quantities are as follows:
$s=x_1 x_2 S$,
$t-m_3^2=x_1(T-m_3^2)$,
$u-m_3^2=x_2(U-m_3^2)$.
Then 
\beqa
\frac{S_4}{S}&=&\frac{s_4}{s}-(1-x_1)\frac{(u-m_4^2)}{s}
-(1-x_2)\frac{(t-m_4^2)}{s}+{\cal O}((1-x_1)(1-x_2)) \,.
\eeqa

If we denote by $\sigma$ the hadron-level cross section and by 
${\hat \sigma}$ the parton-level cross section, then we have the relations 
\beqa
\sigma&=&\int dx_1 \, dx_2 \, \phi(x_1) \, \phi(x_2) \,  {\hat \sigma} 
\nonumber \\
\Rightarrow \frac{d\sigma}{dT\, dU}&=& \int dx_1 \, dx_2 \, 
\frac{dt}{dT} \, \frac{du}{dU} \,  
\phi(x_1) \, \phi(x_2) \, \frac{d{\hat \sigma}}{dt \, du} 
\eeqa
where $\phi(x)$ are the parton distributions.
Now $dt=x_1 dT$ and $du=x_2 dU$, so 
\beqa
\frac{d\sigma}{dT\, dU}&=& \int dx_1 \, dx_2 \, x_1 \, x_2 \, 
\phi(x_1) \, \phi(x_2) \, \frac{d{\hat \sigma}}{dt \, du} \nonumber \\
\Rightarrow \sigma&=& \int dT \, dU \, dx_2 \, ds_4 \,  
\frac{x_1 x_2}{x_2 S+T-m_3^2} \, \phi(x_1) \, \phi(x_2) \, 
\frac{d{\hat \sigma}}{dt \, du} \, ,
\eeqa
where we used the relation
\beq
x_1=\frac{s_4-m_3^2+m_4^2-x_2(U-m_3^2)}{x_2 S+T-m_3^2} \, .
\label{x1}
\eeq

Writing explicitly the kinematical limits on the integrals, we get the 
precise expression for the hadronic cross section 
\beqa
\sigma_{p_{h1}p_{h2} \rightarrow p_3 p_4}(S)&=&
\int_{T_{min}}^{T_{max}} dT 
\int_{U_{min}}^{U_{max}} dU 
\int_{x_{2min}}^1 dx_2 
\int_0^{s_{4max}} ds_4
\nonumber \\ &&
\times \frac{x_1 x_2}{x_2 S+T-m_3^2} \,
\phi(x_1) \, \phi(x_2) \, 
\frac{d^2{\hat\sigma}_{p_1 p_2 \rightarrow p_3 p_4}}{dt \, du}
\eeqa
where $x_1$ is given in Eq. (\ref{x1}), and the limits on the integrals are 
\beq
T_{^{max}_{min}}=-\frac{1}{2}(S-m_3^2-m_4^2) \pm 
\frac{1}{2} \sqrt{(S-m_3^2-m_4^2)^2-4m_3^2 m_4^2} \, ,
\eeq
\beq
U_{max}=m_3^2+\frac{m_3^2 S}{T-m_3^2} \, , \quad U_{min}=-S-T+m_3^2+m_4^2 \, ,
\eeq
\beq
x_{2min}=\frac{m_4^2-T}{S+U-m_3^2}\, ,
\eeq
and $s_{4max}=x_2(S+U-m_3^2)+T-m_4^2$.

The Born-level differential partonic cross section is written as
\beq
\frac{d^2{\hat\sigma}^B_{p_1p_2\rightarrow p_3p_4}}{dtdu}
=F^B_{p_1p_2 \rightarrow p_3p_4} \delta(s_4)=
\frac{\mid {\cal M}_{p_1 p_2 p_3 p_4} \mid^2}
{16\pi s^2}\delta(s_4) \, ,
\label{fb}
\eeq
where the amplitude squared has to be specified for the partonic process 
studied.

For the $t$-channel process $ub \rightarrow dt$ we have 
\beq
|{\cal M}_{ub \rightarrow dt}|^2= 
\frac{4 \pi^2 \alpha^2 V_{tb}^2 V_{ud}^2}{\sin^4\theta_W}
\frac{s(s-m_t^2)}{(t-m_W^2)^2} \, .
\eeq

For the $t$-channel process ${\bar d}b \rightarrow {\bar u} t$ we have 
\beq
|{\cal M}_{{\bar d}b \rightarrow {\bar u}t}|^2= 
\frac{4 \pi^2 \alpha^2 V_{tb}^2 V_{ud}^2}{\sin^4\theta_W (t-m_W^2)^2}
\left[(s+t)^2-(s+t)m_t^2\right] \, .
\eeq

For the $s$-channel process $u {\bar d} \rightarrow {\bar b} t$ we have 
\beq
|{\cal M}_{u {\bar d} \rightarrow {\bar b} t}|^2= 
\frac{4 \pi^2 \alpha^2 V_{tb}^2 V_{ud}^2}{\sin^4\theta_W} 
\frac{t(t-m_t^2)}{(s-m_W^2)^2} \, .
\eeq

For the associated production process $bg \rightarrow tW^-$ we have 
\beq
|{\cal M}_{bg \rightarrow tW^-}|^2= 
\frac{4 \pi^2 V_{tb}^2 \alpha_s \alpha}{3 m_W^2 \sin^2\theta_W}
\left(\frac{A_1}{u_1^2}-\frac{2 A_2}{u_1 s}+\frac{2 A_3}{s^2}\right) \, ,
\eeq
where
$A_1=-u_2(s-m_t^2-m_W^2)(m_W^2+m_t^2/2)
-(t_1/2)(-2m_W^4+m_W^2 m_t^2+m_t^4)
-2 u_2 m_t^2 (2 m_W^2+m_t^2)$,
$A_2=-t_1(-m_W^2+m_t^2)m_W^2-(u_2/2) t_1 m_t^2
-u_1 (u_2/2) m_t^2-s m_W^2 m_t^2-(s/2) m_t^4$,
$A_3=-u_1 (s/4) (2 m_W^2+m_t^2)$
where $\alpha=e^2/(4\pi)$ and $\alpha_s$ is the strong coupling.

The expressions for the corresponding processes involving charm quarks and 
Cabibbo-suppressed contributions are similar.

The electroweak parameters used in the above expressions and 
throughout the paper are the following \cite{PDG}:

$m_W=80.403$ GeV, 

$G_F=\frac{g_W^2}{4\sqrt{2}m_W^2}=1.16637 \times 10^{-5}$ GeV$^{-2}$,

$\sin^2\theta_W|_Z=0.23122$,

$e=g_W \sin\theta_W=0.31404$,

$\alpha=e^2/(4\pi)=7.848\times 10^{-3}$.

\vspace{2mm}

For the CKM matrix elements we have used \cite{PDG}

$V_{ud}=0.97383$, $V_{us}=0.2272$, $V_{ub}=0.00396$,

$V_{cd}=0.2271$, $V_{cs}=0.97296$, $V_{cb}=0.04221$,

$V_{td}=0.00814$, $V_{ts}=0.04161$, $V_{tb}=0.9991$.

\end{document}